\documentclass[a4paper,11pt]{article}
\pdfoutput=1 % if your are submitting a pdflatex (i.e. if you have
\usepackage{jcappub} % for details on the use of the package, please
\usepackage{aas_macros}
\usepackage{graphicx}	% Including figure files
\usepackage{amsmath,amssymb,amsfonts,mathrsfs} % Advanced maths commands
\usepackage{bm}		% Bold maths symbols, including upright Greek
\usepackage{soul}
\usepackage[flushleft]{threeparttable}
\usepackage{multirow}
\usepackage{xspace}
\usepackage{Custom_Functions}

\usepackage[normalem]{ulem}
\usepackage[T1]{fontenc}

\DeclareUnicodeCharacter{0303}{$\alpha$} % Fixes minus sign issue
\defcitealias{puchwein2015}{P15}		% Frequently used References
\defcitealias{puchwein2019}{P19}		% Frequently used References
\defcitealias{haardt2012}{HM12}		% Frequently used References
\defcitealias{khaire2015a}{KS15}		% Frequently used References
\defcitealias{khaire2019a}{KS19}		% Frequently used References
\defcitealias{onorbe2017}{OH17}		% Frequently used References
\defcitealias{faucher2020}{FG20}		% Frequently used References
\defcitealias{gaikwad2018}{G18}	% Frequently used References

\usepackage[usenames, dvipsnames]{color}
\definecolor{mycolor}{RGB}{0,128,0}
\definecolor{newaddcolor}{RGB}{255,0,255}

\usepackage[normalem]{ulem}

\usepackage{xr}
\title{Measuring photo-ionization rate and mean free path of He~{\sc ii} ionizing photons at \zrange{2.5}{3.6}: Evidence for late and rapid HeII reionization Part-II}
\author[1,a,b]{Prakash Gaikwad}\note{Corresponding author}
\author[b]{Fredrick B. Davies}
\author[c]{and Martin G. Haehnelt}
\affiliation[a]{Department of Astronomy, Astrophysics and Space Engineering, Indian Institute of Technology Indore, Simrol, MP 453552, India}
\affiliation[b]{Max-Planck-Institut f\"{u}r Astronomie, K\"{o}nigstuhl 17, D-69117 Heidelberg, Germany}
\affiliation[c]{Kavli Institute for Cosmology and Institute of Astronomy, Madingley Road, Cambridge, CB3 0HA, UK}
\emailAdd{gaikwad@iiti.ac.in}
\emailAdd{davies@mpia.de}
\emailAdd{haehnelt@ast.cam.ac.uk }
\abstract{
We present measurements of the spatially averaged He~{\sc ii} photo-ionization
rate ($\langle \Gamma_{\rm HeII} \rangle$), mean free path of He~{\sc ii}
ionizing photons ($\lambda_{\rm mfp, HeII}$), and \HeII fraction ($f_{\rm
HeII}$) across seven redshift bins within the redshift range $2<z<4$. The measurements
are obtained by comparing the observed effective optical depth distribution of
\HeII ($\tau_{\rm eff, HeII}$) with models generated by post-processing of the
Sherwood simulation suite using our code \excitecode. With \excitecode, we
efficiently explore a large parameter space ($\sim 15000$ models) by varying $\lambda_{\rm mfp, HeII}$ and $\langle \Gamma_{\rm HeII}
\rangle$. We employ a non-parametric Anderson-Darling test for the cumulative
distribution of $\tau_{\rm eff, HeII}$ to simultaneously measure
$\lambda_{\rm mfp, HeII}$ and $\langle \Gamma_{\rm HeII} \rangle$.  Our
measurements account for possible observational and modeling uncertainties
stemming mainly  from the finite signal-to-noise ratio of the observed data and thermal
parameter uncertainties.  We find significant evolution, with the best-fit
$\langle \Gamma_{\rm HeII} \rangle$ and $\lambda_{\rm mfp, HeII}$ decreasing by
factors of $\sim 4.32$ and $ \sim 3.27$, respectively, from $z = 2.88$ to $z =
3.16$. This decreasing trend in $\langle \Gamma_{\rm HeII} \rangle$ and
$\lambda_{\rm mfp, HeII}$ suggests evolution of the size of ionized regions
implying that HeII reionization is still ongoing at these redshifts.
Based on these measurements, we constrain the emissivity at the \HeII ionization
frequency ($\epsilon_{228}$) and \HeII ionizing photon emission rate
($\dot{n}$), finding consistency with results from galaxy and QSO surveys.
Comparison of our measured parameters with widely used uniform UVB models
supports a scenario where He~{\sc ii} reionization is not completed before
$z\sim2.74$.  Our measured  evolution is  complementary and    in good agreement
with recent measurements of thermal parameters of the IGM, suggesting a coherent picture of
rather late and rapid \HeII reionization.
}
\begin{document}
%\label{firstpage}
\date{}
\maketitle
\flushbottom

%%%%%%%%%%%%%%%%%%%%%%%%%%%%%%%%%%%%%%%%%%%%%%%%%

%%%%%%%%%%%%%%%%%%%%%%%%%%%%%%%%%%%%%%%%%%%%%%%%%%%%%%%%%%%%%%%%%%%%%%%%%%%%
%%%%%%%% THIS ABSTRACT DOES NOT CONTAIN ALIASES USED IN THE TEX FILE %%%%%%%
%%%%%%%%%%%%%%%%%%%%%%%%%%%%%%%%%%%%%%%%%%%%%%%%%%%%%%%%%%%%%%%%%%%%%%%%%%%%
%%%%%%%%%%%%%%%%%%%%%%%%%%%%%%%%%%%%%%%%%%%%%%%%%

\section{Introduction}
\label{sec:introduction}
%%%%%
Hard photons ($E_{\nu} > 54.4$ eV) emitted by Quasi-Stellar Objects (QSOs) play
a crucial role in the second major phase transition of the IGM, the ionization
of \HeII to \HeIII \citep{picard1993,jakobsen1994,reimers1997,madau1999,
kriss2001,zheng2004,shull2004b,furlanetto2008b,mcquinn2009}.  Observations
suggest that \HeII reionization occurs within the redshift range $2<z<5$ due to
an increasing contribution of QSOs to the total ionizing emissivity
supplementing the contribution from massive stars in galaxies
\citep{haardt2012,khaire2015a,khaire2017,puchwein2019}.  Understanding \HeII
reionization is not only important in its own right but it also provides more
general insights into reionization, potentially also applicable to \HI
reionization  \citep[see the review article by][]{mcquinn2015b}.  With the large number
of unprecedented quality QSO absorption spectra obtained in the last two
decades, it is now possible to study the process of \HeII reionization
$(2<z<5)$ in detail \citep{worseck2011,omeara2017,murphy2019}.

%%%%%
The process of \HeII reionization can be constrained through several key
parameters that describe the thermal and ionization state of the IGM. These
parameters include the normalization ($T_0$) and slope ($\gamma$) of the
temperature-density relation (TDR), the \HeII fraction (\fHeII), the mean free
path of \HeII ionizing photons (\LmfpHeII), and the \HeII photo-ionization rate
(\GHeII). The thermal parameters ($T_0$ and $\gamma$) are sensitive to the
amount and manner of energy deposition into the IGM, which is  influenced by the
spectral energy distribution of QSOs and the timing of reionization
\citep{hui1997,mcquinn2015a,gaikwad2019}.  Conversely, \fHeII, \LmfpHeII, and
\GHeII depend significantly on the nature of the reionization process,
particularly the patchiness of reionization \citep{shull2004, furlanetto2008a,
dixon2014, mcquinn2014, davies2014, davies2017,basu2024}. Robust measurement of
these IGM parameters has been one of the primary goals of recent studies. The
thermal parameters  have been measured from  increasingly larger observed
samples of \HI \lya forest data \citep{schaye2000,lidz2010,becker2011,boera2014,
rorai2017b,hiss2018,telikova2019,walther2019,gaikwad2020,ondro2021,gaikwad2021,
ondro2023}.  \cite{gaikwad2021} in particular provide a consistent measurement
of the   evolution of thermal parameters from a variety of  \lya flux statistics
that suggests that  \HeII reionization was late and rapid.
%%%%%%

Here we constrain the evolution of the \fHeII, \GHeII, and \LmfpHeII parameters
using the \HeII \lya forest, complementing the thermal parameter measurements
from the \HI \lya forest.  Measuring these parameters presents significant
challenges, primarily due to the difficulty of observations of the \HeII \lya
forest, which requires space-based telescopes
\citep{heap2000,reimers2005,fechner2006,syphers2014}.  Suitable unobscured
sightlines for \HeII \lya forest observations are scarce, as background QSO must
be sufficiently bright and free from contamination by high \HI column density
foreground systems \citep{worseck2016,makan2021}.  Previous studies have
constrained \GHeII and \fHeII using uniform UVB simulations, neglecting  spatial
fluctuations in \GHeII \citep[but see][]{davies2017}. Notably, while constraints
exist for \GHeII and \fHeII, there are currently no observational constraints
available for \LmfpHeII.  

The traditional method of measuring the mean free path is very difficult  for
\HeII ionizing photons as it requires to observe the flux at the \HeII Lyman
continuum ($\lambda_{\rm LyC, HeII} \sim 228$ \text{\AA}) \citep[see similar
approaches by][for \HI
reionization]{prochaska2009,fumagalli2013,worseck2014,becker2021,zhu2023}.
However, an alternative approach is possible that involves modeling the
effective optical depth distribution of \HeII with models of patchy
reionization, where the mean free path is varied alongside \GHeII and \fHeII.  A
similar  methodology has been demonstrated to be successful in the context of \HI
reionization \citep{gaikwad2023,davies2024}. Modeling fluctuations in the
ionizing radiation field is crucial for this approach. The validity of previous
\fHeII and \GHeII measurements were generally limited due to a lack of modeling
these fluctuations.

Modeling the patchy inhomogeneous \HeII reionization process presents another
significant challenge due to the need for high dynamic range simulations
\citep{davies2014,leplante2016,leplante2017a,leplante2018,kapahtia2024,basu2024}.
To adequately capture the patchy nature of reionization requires a balance
between the presence of massive halos capable of hosting QSOs and the necessary
resolution of the simulated \HeII \lya forest spectra to match
observations \citep{worseck2019}.  Early in the \HeII reionization process,
patchiness is expected due to the overlapping ionized bubbles, gradually
leading to a more uniform ionizing radiation field.  This is further complicated
by the fact that the ionizing sources are expected to be short-lived.
Simulating this is computationally expensive, and widely employed  radiative
transfer simulations using the M1 closure condition may not accurately capture
the radiation field, particularly in the presence of bright QSOs in clustered
environments \citep{aubert2008,garaldi2022}.  Anisotropic emission from QSOs,
with the possibility of intersecting beams, further complicates the modeling
\citep{rosdahl2013}. Thus, alternative methods are necessary for accurately
capturing the radiation field geometry during \HeII reionization.

In this paper, we model the \HeII ionizing radiation field using a ray-tracing
approach as an alternative to the M1-based method \citep[see][for other ray-tracing
codes]{abel2002,trac2007,cantalupo2011,hartley2019,cain2024}.  Traditional ray-tracing approaches are
computationally demanding due to the complexity of casting rays, scaling with
the number of sources as $\mathcal{O}(N^2)$ \citep[e.g., {\sc c$^2$ray,
}][]{mellema2006}. However, our recently developed ray-tracing code,
\excitefullform (\excitecode), employs efficient octree decomposition, enabling
more favourable scaling with  $\mathcal{O}(N \: \log N)$ \citep{gaikwad2023}.
This advancement allows us to generate large-scale models of patchy
reionization, facilitating measurements of the relevant parameters \citep{wu2021}.  By
combining a large number of simulations with high-quality observations of the \HeII
effective optical depth from \citep{worseck2019}, we constrain the key
parameters governing \HeII reionization (\fHeII, \GHeII and \LmfpHeII).
Constraining these parameters allows the comparison of different models,
providing valuable insights into the complex process of \HeII reionization.

The paper is structured as follows: Section \ref{sec:observation} outlines the
observational data. Sections \ref{sec:simulation} and \ref{sec:method} describe
the hydrodynamical simulations and the theoretical framework employed to model
patchy \HeII reionization, respectively. Section \ref{sec:results} presents the
primary findings of the study. In Section \ref{sec:discussion}, we discuss
current limitations and future improvement needed in this field, while Section
\ref{sec:summary} provides a summary of the work.  Throughout this work, the
comoving and physical distances are prefixed by symbols `c' and `p',
respectively.  We adopt a flat $\Lambda$CDM cosmology consistent with the
parameters reported in \citep{planck2014}: $h=0.678$, $Y=0.24$,
$\Omega_{\lambda} = 0.692$, $\Omega_{\rm m} = 0.308$, $\Omega_{\rm b} = 0.0482$,
$n_{\rm s}=0.961$, $\sigma_{\rm 8} = 0.829$.

%====================================================
\section{Observations}
\label{sec:observation}
%---------------
\InputFig{{HeII_HST_COS_Redshift_Coverage_JCAP}.pdf}{155}%
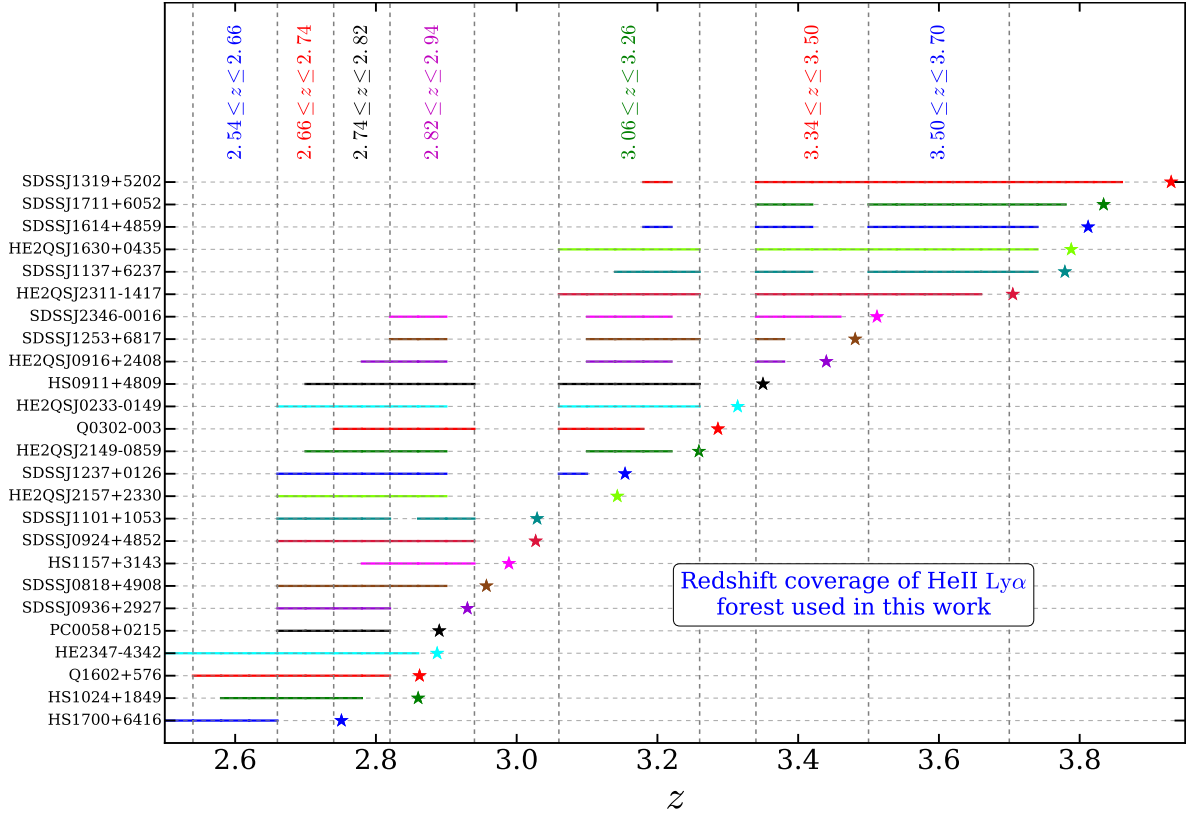
{ The figure shows the redshift coverage of the \HeII \lya forest based on the
sample of 24 quasars used in this work from \cite{worseck2019}. Vertical dashed
lines denote the division into seven redshift bins, each labeled with its
corresponding redshift range.  The figure shows unavailable \taueffHeII values 
due to exclusion of the quasar proximity region, as well as
regions contaminated by geocoronal emission. Mock \HeII \lya forest spectra are
simulated to match observational data properties, including gaps, noise
properties and instrumental broadening, ensuring fair comparison between model
and data (see \S \ref{sec:observation} and \ref{sec:results}).
}{\label{fig:redshift-coverage}}
%---------------
In this work, we use publicly available measurements of \HeII effective optical
depths (\taueffHeII) from \cite{worseck2019}. These measurements are derived
from observations of the \HeII \lya forest along 24 quasar sightlines using the
Cosmic Origins Spectrograph  on board the Hubble Space Telescope (HST-COS).
These sightlines were chosen  to ensure sufficient brightness of the background
quasars to obtain high signal-to-noise ratio (\SNR) spectra, while also avoiding
interference from intervening optically thick \HI systems at low redshifts \citep{worseck2011,worseck2016}.

Our analysis considers the redshift coverage of the \HeII \lya forest along
these sightlines as shown in Figure \ref{fig:redshift-coverage}. Exclusions are
made in certain spectral regions to enhance the reliability of our analysis.
Specifically, segments between the rightmost edge of the spectra and the
emission redshift of the quasar (indicated by star symbols) are excluded due to the 
potential influence from intense radiation in the proximity zone of the
background quasar. Similarly, the spectral regions encompassing $2.94 \leq z
\leq 3.06$ and $3.26 \leq z \leq 3.34$ are omitted from our analysis due to
contamination by geocoronal \HI \lya and O~{\sc i} line emission,
respectively.

The observations primarily employ the G140L grating of HST-COS, with occasional
use of the G130M grating. These gratings correspond to velocity resolutions
(FWHM) of approximately $150$ \kmps and $20$ \kmps, respectively. The
signal-to-noise ratio per pixel across the different sightlines ranges from 3
to 19. It is noteworthy that the noise properties of these spectra exhibit
Poisson-limited conditions.  A maximum likelihood approach based on the
Poisson distribution is adopted to derive \taueffHeII from observations in
\cite{worseck2019}.  This methodology enables the extraction of reliable
measurements for \taueffHeII across all sightlines.  Additionally, a
comprehensive analysis of both observational statistical uncertainties and
systematic uncertainties is carried out to provide a robust estimation of the
measurement uncertainties associated with \taueffHeII. We follow the same procedure
for  deriving the \taueffHeII measurements from our simulations.

The observational dataset is divided into seven distinct redshift bins, $z \pm
\Delta z = 2.60 \pm 0.06,  2.70 \pm 0.04,  2.78 \pm 0.04,  2.88 \pm 0.06, 3.16
\pm 0.10, 3.42 \pm 0.08,$ and $3.60 \pm 0.10$. Within each redshift bin, the
\taueffHeII values are computed using an averaging window of $\Delta z_{\rm
\tau} = 0.04$. This binning strategy is chosen to maximize the count of
\taueffHeII measurements per redshift bin \citep{worseck2019}.  Consequently,
the quantities of \taueffHeII measurements within the aforementioned seven
redshift bins are $N_{\rm \tau_{\rm eff}} = 11, 24, 28, 28, 42, 23,$ and $29$,
respectively. The cumulative distribution functions of \taueffHeII measurements
are calculated and subsequently compared with models to measure the mean free
path and \HeII photo-ionization rate. To facilitate this analysis, we generate
simulated spectra that closely replicate the observed sample regarding the
redshift path length, spectral gaps, signal-to-noise ratio (\SNR), and line
spread function (LSF) characteristics.  For  simulated mock datasets we adopt
binning analogous to that of the observations for our \taueffHeII calculations,
as outlined in Section \ref{sec:results}.

%%====================================================
\section{Simulations}
\label{sec:simulation}
%---------------

\begin{table*}
\centering
\caption{The table gives a summary of \excitecode models performed in this work to simulate \HeII reionization}
\resizebox{\textwidth}{!}{
\begin{tabular}{ccccccl}
\hline  \hline 
Simulation  & ${\rm N_{\rm Grid, \Gamma_{\rm HeII}}}$ & $N_{\rm model}$ & ${\rm M_{\rm cutoff}}$ & $\beta$ & $\zeta$ & Motivation\\
\hline
L160N2048   & 512  & 164 & $10^{11}$ & 1.41 & 2/3  & Default model for parameter estimation \\
\hline 
L160N2048   & 256  & 4   & $10^{11}$ & 1.41 & 2/3  & Convergence of \GHeII/\GHeIIavg maps (right panel of \figref{fig:convergence-test-1}, \ref{fig:convergence-test-2})\\
L160N2048   & 1024 & 4   & $10^{11}$ & 1.41 & 2/3  & Convergence of \GHeII/\GHeIIavg maps (right panel of \figref{fig:convergence-test-1}, \ref{fig:convergence-test-2})\\
\hline 
L40N512     & 512  & 4   & $10^{11}$ & 1.41 & 2/3  & Convergence of box size (left panel of \figref{fig:convergence-test-1}, \ref{fig:convergence-test-2}) \\
L80N1024    & 512  & 4   & $10^{11}$ & 1.41 & 2/3  & Convergence of box size (left panel of \figref{fig:convergence-test-1}, \ref{fig:convergence-test-2}) \\
\hline
L160N512    & 512  & 4   & $10^{11}$ & 1.41 & 2/3 & Convergence of mass resolution (middle panel of \figref{fig:convergence-test-1}, \ref{fig:convergence-test-2}) \\
L160N1024   & 512  & 4   & $10^{11}$ & 1.41 & 2/3 & Convergence of mass resolution (middle panel of \figref{fig:convergence-test-1}, \ref{fig:convergence-test-2}) \\
\hline
L160N2048   & 512  & 4 & $10^{9}$  & 1.41 & 2/3 & Modeling uncertainty: Effect of $M_{\rm cutoff}$ (\figref{fig:parameter-uncertainty-modeling}) \\
L160N2048   & 512  & 4 & $10^{12}$ & 1.41 & 2/3 & Modeling uncertainty: Effect of $M_{\rm cutoff}$ (\figref{fig:parameter-uncertainty-modeling}) \\
\hline
L160N2048   & 512  & 4 & $10^{11}$ & 1.41 & 1/3 & Modeling uncertainty: Effect of $\zeta$ (\figref{fig:parameter-uncertainty-modeling}) \\
L160N2048   & 512  & 4 & $10^{11}$ & 1.41 & 3/4 & Modeling uncertainty: Effect of $\zeta$ (\figref{fig:parameter-uncertainty-modeling}) \\
\hline
L160N2048   & 512  & 4 & $10^{11}$ & 1.0 & 2/3 & Modeling uncertainty: Effect of $\beta$ (\figref{fig:parameter-uncertainty-modeling}) \\
L160N2048   & 512  & 4 & $10^{11}$ & 2.0 & 2/3 & Modeling uncertainty: Effect of $\beta$ (\figref{fig:parameter-uncertainty-modeling}) \\
\hline \hline
\end{tabular}
}
\\
\label{tab:simulations}
\end{table*}

We post-process the \sherwood simulation suite using our code
\excitefullform (\excitecode), designed to model fluctuations in the ionizing
background \citep{gaikwad2023}. The \sherwood suite is performed with the
smoothed particle hydrodynamic code \gthree \citep{springel2005, bolton2017}.
All simulations are performed utilizing a modified version of the uniform
ultraviolet background (UVB) model by \cite{haardt2012}. To ensure consistency
with the thermal parameter evolution measured by \cite{becker2011}, the \HeII
photo-heating rates are increased by 20 percent. We use density, velocity
fields, and halo catalogs from the \sherwood simulation.  The temperature and \HeII
fraction fields from the simulation output are intentionally excluded, as these
fields are modeled during the post-processing step of our analysis using
\excitecode.

The simulations employ simplified star-formation criteria  of $T<10^5$ and
$\Delta>1000$, while detailed galaxy and quasar formation is not explicitly
tracked \citep{viel2005}. Instead, halos are identified using an on-the-fly halo finder
algorithm, and their positions are used to choose  the locations of ionizing
sources (QSOs) in the simulations. The \sherwood simulation suite comprises
varying box sizes, ranging from $10 \: h^{-1} \: {\rm cMpc}$ to $160 \: h^{-1} \:
{\rm cMpc}$, with particle numbers spanning $512^3$ to $2048^3$. A summary of
the utilized \sherwood simulations can be found in Table \ref{tab:simulations}.

The observations of the \HeII \lya forest are obtained at a moderate resolution of
approximately $150$ \kmps.  For the analysis our primary focus centers on the L160N2048
simulation, offering the broadest dynamic range within the \sherwood suite,
which allows us to explore parameter variations that model fluctuations in the
ionizing radiation field.  Additionally, we utilize the L40N512, L80N1024,
L160N512, and L160N1024 models to conduct convergence tests related to box size
and mass resolution (see appendix \ref{app:convergence-test}).

\section{Method}
\label{sec:method}
Modeling the flucutuations in the ionizing radiation field caused by QSOs is
crucial when comparing statistical properties of the \HeII
\lya forest in simulations and observations.  In our previous  work
\citep{gaikwad2023}, we introduced a tool named \excitecode for modeling these
fluctuations during \HI reionization. In the subsequent section, we outline the
primary steps involved in modeling patchy \HeII reionization and subsequently
discuss our approach to forward-modeling simulated \HeII \lya forest spectra.

\subsection{Brief overview of \excitecode}
\label{subsec:excite}
\InputFigCombine{{MFP_Comparison}.pdf}{155}%
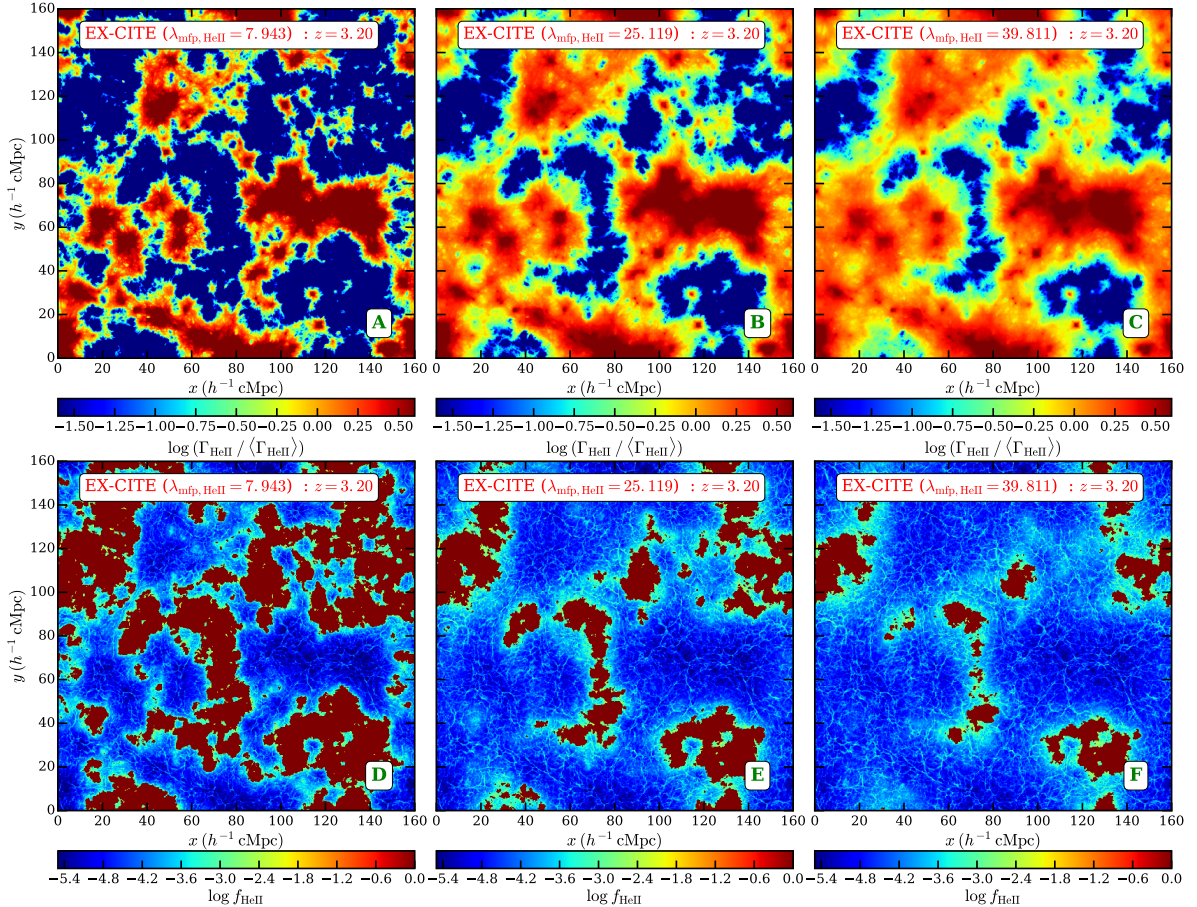
{ Panel A, B and C show fluctuations in the \HeII photo-ionization rate
    $(\Gamma_{\rm HeII} / \langle \Gamma_{\rm HeII} \rangle)$ from \excitecode
    for different values of true mean free path \LmfpHeII $ = 7.943, 25.119 $
    and $39.811$ (in $h^{-1} \: {\rm cMpc}$), respectively.  Panel D, E and F are
    similar to panels A, B and C except that maps of \HeII fractions are shown.
    The maps are produced on $512^3$ grids and the thickness of all slices are
    the same. Note that the simulation have a box size $L=160 \: h^{-1} \: {\rm
    cMpc}$ and number of particles $N_{\rm particle} = 2048^3$. The color schemes of
    panel A, B, C (and also D, E, F) are identical  for a fair
    comparison. With the increase in mean free path (panel A, B and C), ionizing
    radiation spreads to larger distances. This is accompanied by a decrease in
    neutral fraction as the radiation field ionizes the IGM at larger distances
    (panel D, E and F).  In the bottom panels, we fix  $\langle \Gamma_{\rm
    HeII} \rangle = 10^{-15} \;\; {\rm s^{-1}}$.
}{\label{fig:mfp-slice-comparison}}
To model the patchy nature of inhomogeneous \HI and \HeII reionization, we have developed
\excitecode, as described in \cite{gaikwad2023}. In this framework, we
start by identifying halos and assigning normalized emissivity weights to
each halo with mass $M_{\rm halo}$ as,
\begin{equation}{\label{eq:emissivity-weights}}
\begin{aligned}
    w_{\rm halo, k} &=  M^{\beta}_{\rm halo, k} \;\; / \;  \sum \limits_{k=1}^{N_{\rm halo}} M^{\beta}_{\rm halo, k}  &\; {\rm for} \; &M_{\rm halo} \geq M_{\rm cutoff}, \\
    &=  0 &\; {\rm for} \; &M_{\rm halo} < M_{\rm cutoff}.
\end{aligned}
\end{equation}
The choice of employing emissivity weights rather than absolute emissivities is because this enables the modeling of fluctuations in \HeII
photo-ionization rate independently from the spatially averaged photo-ionization
rate, allowing efficient  exploration of the parameter space.  The parameters of halo
mass cutoff ($M_{\rm cutoff}$) and emissivity-halo mass index ($\beta$) dictate
both the allocation of emissivity to halos and the dependence of the emissivity
on the mass of the halos.  The specific values for $M_{\rm cutoff}$ and $\beta$
are given in Table \ref{tab:simulations}. Our default models adopt $M_{\rm
cutoff} = 10^{11} \: {\rm M_{\odot}}$ and $\beta = 1.41$ \citep{finlator2011,
sanderbeck2020}.

With the assigned emissivity weights for each halo, the fluctuations in the
\HeII photo-ionization rate at cell $i$ can be determined by adding
contributions from all halos at positions $k$ and accounting for the influence
of IGM attenuation along the sightline due to the finite mean free path
$\lambda(x)$ as,
\begin{equation}{\label{eq:photo-ionization-rate}}
\begin{aligned}
    \frac{\Gamma_{{\rm HeII}, i}}{\langle \Gamma_{{\rm HeII}} \rangle} &= f_{\rm norm} \: \sum \limits_{i=1, i \neq k}^{N_{\rm source}} \frac{w_{{\rm halo},k}}{(4 \: \pi \: r_{ik})^2} \; \exp \bigg[ - \int \limits_{r_i}^{r_k} \frac{dx}{\lambda(x)} \bigg],
\end{aligned}
\end{equation}
where $f_{\rm norm}$ is a normalization constant that ensures that the average of \GHeII
/\GHeIIavg over the simulation volume is normalized to unity.  In the \excitecode
formalism, the mean free path $\lambda(x)$ at any given cell depends on the
local photo-ionization rate fluctuations as,
\begin{equation}{\label{eq:fluctuating-mfp}}
\begin{aligned}
    \lambda(x) &=  \lambda_{0} \; \Delta^{\xi} \; \bigg[ \frac{\Gamma_{\rm HeII}(x)}{\langle \Gamma_{\rm HeII} \rangle} \bigg]^{\zeta} \; \bigg[ \frac{E_{\rm bin}}{E_{\rm ion,HeII}} \bigg]^{0.9},
\end{aligned}
\end{equation}
where \Lmfp represents the spatially averaged mean free path parameter, $\Delta$
is the local overdensity, $E_{\rm ion, \HeII}=54.4 \: {\rm eV}$ denotes the
ionization potential of \HeII, $E_{\rm bin}$ is  the energy of photons in a
specific frequency bin and $\xi$ characterizes the dependence of the mean free path on the local overdensity \citep{davies2016,gaikwad2023,davies2024}. We
find that $\xi=0.5-1.5$ produces results consistent with  radiative transfer
simulations \citep{munoz2016}. In this study, we adopt a mono-frequency approach
with $E_{\rm bin}=60.2 \; {\rm eV}$ and $\xi=1.5$. Our default models assume
$\zeta=2/3$, though we also investigate the range $1/3$ to $3/4$ for $\zeta$ to
assess modeling uncertainties (refer to Table \ref{tab:simulations}). The 
value of $\zeta$ can also be predicted from the \HI column density distribution function with 
slab modeling of the IGM \cite{faucher2009,haardt2012}. Eq. \ref{eq:fluctuating-mfp} ignores  that the photons with energy $>54.4$ eV that ionize \HeII can also be absorbed by HI. However, this 
effect is less significant because the \HI photo-ionization cross-section drops quickly below $13.6$ eV ($\propto \; \nu^{-3}$) and the mean free path of \HI ionizing photon is very large \citep[\HI reionization completes by $\sim 5.2$ as shown in][]{gaikwad2023}.
n this work, we evaluate the mean free path at the \HeII ionizing edge using the mono-frequency implementation of \excitecode, which, while neglecting the frequency dependence introduced by the spectral and column density distribution shape, enables an efficient exploration of a large parameter space. This simplification may lead to a modest underestimation of the uncertainties in the mean free path; however, as we show later, the dominant contribution to the total uncertainty arises from the thermal parameters of the IGM, making the mono-frequency treatment a reasonable assumption. The
photo-ionization rate fluctuations, as defined in Eq.
\ref{eq:photo-ionization-rate}, are independent of the spectral energy
distribution (SED) of the sources. However, the source SED impacts photo-heating
rates and subsequently the gas temperature of the ionizing cells.  Rather than
changing the QSO SED, we vary the temperature using data from \cite[][see \S
for details]{gaikwad2021}. Within \excitecode, we employ octree summation to
compute the contribution of all sources (as given by Eq.
\ref{eq:photo-ionization-rate}). This octree approach enables computation of the
photo-ionization rate fluctuations field with $\mathcal{O}(N \: \log N)$
operations, resulting in the efficiency required to explore an extensive
parameter space with a high-resolution \GHeII field \citep[see ][for details of
the numerical implementation]{gaikwad2023}.

Using the fluctuations in photo-ionization rate, the \HeII fraction
$(f_{\rm HeII})$ at each location can be calculated using,
\begin{equation}%
{\label{eq:nHeII-ex-cite}}
    f_{\rm HeII} = \frac{\mu_{e} \: n_{\rm He} \: \alpha_{\rm HeII}(T)}{\langle \Gamma_{\rm HeII} \rangle \times [\Gamma_{\rm HeII} / \langle \Gamma_{\rm HeII} \rangle]_{\rm EX-CITE}},
\end{equation}
where $n_{\rm He}$ is the number density of helium, $\mu_{e}= \big[ (1-Y) \:
f_{\rm HII} + Y/4 \: (f_{\rm HeII} + 2 \: f_{\rm HeIII}) \big] / (1-Y)$ denotes
the mean molecular weight of electrons, $\alpha_{\rm HeII}(T)$ is the \HeII
recombination rate coefficient,   $[\Gamma_{\rm HeII} / \langle \Gamma_{\rm
HeII} \rangle]_{\rm EX-CITE}$ are the fluctuations of the photo-ionization rate as
given in Eq. \ref{eq:photo-ionization-rate} and \GHeIIavg is the spatially
averaged \HeII photo-ionization rate, a free parameter in our analysis. Eq.
\ref{eq:nHeII-ex-cite}  assumes  photo-ionization equilibrium, which generally
holds true in ionized and neutral regions.\footnote{Note that we call here
regions where HeII is not yet ionized to HeIII neutral, in analogy to the
reionization of hydrogen.} However, this assumption can yield unrealistic
$f_{\rm HeII}$ values exceeding 1. To address this, cells are treated as neutral
when $\Gamma^{-1}_{\rm HeII} > t_{\rm Hubble}$ where $t_{\rm Hubble}$ is the
Hubble time Although photo-ionization equilibrium might not hold at ionization
fronts, where the photo-ionization rate changes quickly, these regions occupy a
small volume compared to neutral or ionized regions. Consequently, for practical
considerations, the assumption of photo-ionization equilibrium is typically
sufficient.

The parameter \Lmfp, as defined in Eq. \ref{eq:fluctuating-mfp}, is different
from the conventional definition of the mean free path (hereafter true mean
free path, \LmfpHeII) typically measured in observations. \Lmfp is a global
parameter  that does not explicitly depend  on the neutral (\HeII) fraction. It
serves as a practical parameter for modeling the photo-ionization rate
fluctuations, and depends on the density distribution, as well as location and
properties of the ionizing  sources.  
The local mean free path in each cell is then computed by scaling 
$\lambda_0$ using the local values of density and $\Gamma_{\rm HeII}$, 
as described in Equation \ref{eq:fluctuating-mfp}. Following \cite{gaikwad2023}, our
measurements directly constrain the physical  mean free path \LmfpHeII instead
of the model parameter \Lmfp. For the determination of  \LmfpHeII, we calculate
the \HeII Lyman continuum optical depth using all skewers ($2048^3$) oriented
along any axis ($x$, $y$, or $z$) within our simulation box as,
\begin{equation}%
{\label{eq:tau-lyc}}
\tau_{\rm Lyc, HeII } = \int f_{\rm HeII} \;\; n_{\rm He} \;\; \sigma_{\rm HeII, ion} \;\; dx, 
\end{equation}
where $n_{\rm He}$ is the He number density and
$\sigma_{\rm HeII, ion}=1.588 \times 10^{-18} \; {\rm cm^2}$ is the \HeII
photo-ionization cross-section \citep{verner1994}. 

The integration in above equation is performed through cumulative summation.
Subsequently, the \HeII Lyman continuum flux is evaluated as $F_{\rm Lyc} =
e^{-\tau_{\rm Lyc}}$. Following this, the mean Lyman continuum profile $\langle
F_{\rm Lyc} \rangle$ is obtained by averaging all $2048^3$ $F_{\rm Lyc}$
profiles. For the determination of \LmfpHeII, the mean Lyman continuum
profile is fitted using an exponential function as,
\begin{equation}%
{\label{eq:mfp-rt-simulation}}
\langle F_{\rm Lyc, HeII} \rangle = F_0 \; \exp \bigg[- \frac{x}{\lambda_{\rm mfp, HeII}} \bigg] ,
\end{equation}
where $x$ denotes the distance  (in cMpc) along the sightline.  The exponential
expression above is fitted using two independent parameters: (i) the
normalization $F_0$, and (ii) the true mean free path \LmfpHeII. The \LmfpHeII
as  defined above depends on the \HeII fraction, resembling the approach adopted
for mean free path measurements in observations. In our analysis, \LmfpHeII is
explicitly calculated for each model.  Consequently, both \LmfpHeII and
\GHeIIavg are  the two distinct free parameters that we aim to measure from
observations of the distribution of \taueffHeII.
\figref{fig:mfp-slice-comparison} illustrates the effect of variations in mean
free path on the fluctuations in \HeII photo-ionization rate (\GHeII) and \HeII
neutral fraction (\fHeII). For smaller mean free path, the fluctuations in
\GHeII are more prominent while the fluctuations wane with increasing mean free
path. Consequently the fluctuations in \fHeII closely follows the fluctuations
in \GHeII. This in turn affects the distribution of \taueffHeII seen in
observations. 

The parameters in Eq. \ref{eq:fluctuating-mfp} such as $\lambda_0$ and $E_{\rm bin}$, $\zeta$ and $\xi$ all influence the spatial structure of the photo-ionization rate fluctuations, $\Gamma_{\rm HeII}(x) / \langle \Gamma_{\rm HeII} \rangle$. However, our model is designed to constrain this fluctuation field and not the individual input parameters. 
To clarify this suppose two models, with different combinations of these parameters, produce a similar fluctuation field $\Gamma_{\rm HeII}(x)/\langle \Gamma_{\rm HeII} \rangle$. Both models will then yield  similar predictions for $f_{\rm HeII}$ and $\tau_{\rm eff,HeII}$, and ultimately match the observed data in the same way. This is because we normalize the ionizing background field with an average value $\langle \Gamma_{\rm HeII} \rangle$ before calculating observables. Since the final mean free path we report ($\lambda_{\rm mfp,HeII}$) is derived from the spatial structure of $f_{\rm HeII}$ in the model that best matches the data. Note, however, that the estimated mean free path is little affected by the internal degeneracies in the model parameters.

This situation is conceptually similar to Principal Component Analysis (PCA), where different combinations of parameters can project onto the same principal direction that governs the variation in observables. In our case, even if the underlying internal model parameters ($\lambda_0, E_{\rm bin}, \zeta, \xi$) differ, as long as the resulting fluctuation field is preserved, the predicted observables ($\tau_{\rm eff, HeII}$) and inferred mean free path ($\lambda_{\rm mfp, HeII}$) remain unchanged. 

We emphasize that our primary goal is to constrain the mean free path  ($\lambda_{\rm mfp, HeII}$) through its impact on $f_{\rm HeII}$, rather than to constrain $\lambda_0$ or its associated parameters directly.  As we show later that the dominant source of modeling uncertainty in our analysis arises from the thermal parameters, which play a much more significant role in determining the ionization state of the gas.

\subsection{Parameter variation and model generation}
\label{subsec:parameter-variation}
We systematically vary the mean free path parameter \Lmfp in logarithmic units
(i.e., \logLmfp), ranging from -1.0 to 3.0 in increments of 0.1. This procedure
results in the creation of 41 models characterizing the \GHeII fluctuations at a
single redshift. Utilizing  \sherwood simulation snapshots at four distinct
redshifts ($z=2.4, 2.8, 3.2, 3.6$), we generate a total of 164 \GHeII
fluctuation models with \excitecode, as summarized in Table
\ref{tab:simulations}. While the \GHeII fields are produced at a resolution of
$512^3$, density and velocity fields are generated at a resolution of $2048^3$. Linear
interpolation is employed to map the \GHeII fields from $512^3$ to $2048^3$
grids. The \GHeII fields exhibit satisfactory convergence at a resolution of
$512^3$ (see appendix \ref{app:convergence-test}). 

For  each \GHeII fluctuation model, we also vary the spatially averaged
\GHeIIavg (expressed in units of $10^{-12} \; {\rm s^{-1}}$) in logarithmic
units (i.e., \logGHeIIavg), spanning from -6.0 to -1.5 in increments of 0.05.
This results in the creation of 91 distinct models for each \Lmfp value.
Consequently, a total of $41 \times 91 = 3731$ models are generated at a given
redshift.  To derive the \fHeII field from the \GHeII fluctuations field,
information about the temperature at each location is required. However, due to
\excitecode's lack of time evolution, temperature fluctuations are not
self-consistently modeled.

For the computation of the temperature field, a two-zone model is employed,
following the description in \cite{gaikwad2023}. The regions not yet ionized to
HeIII are attributed to areas with $\Gamma_{\rm 12,HeII} < 10^{-2.6}$ \cite{daloisio2020,gaikwad2023};
otherwise, regions are treated as ionized \footnote{Note that the two conditions on $\Gamma_{\rm HeII}$ are used for different purposes and are therefore not expected to match. In Section \S \ref{subsec:excite}, the condition $\Gamma_{\rm HeII}^{-1} < t_{\rm Hubble}$ is used to determine whether the gas is effectively neutral when computing the ionization state using Equation \ref{eq:nHeII-ex-cite}. In regions with extremely low photo-ionization rates, this condition avoids unphysical values of $f_{\rm HeII} > 1$, which can otherwise arise due to numerical limitations of the ionization equilibrium equation. In contrast, the threshold of $\Gamma_{\rm HeII} < 10^{-2.6}$ used in Section \S\ref{subsec:parameter-variation} is applied when assigning temperatures to cells. This criterion ensures that only highly ionized regions receive a temperature boost from the temperature-density relation. Cells that are partially ionized, or still predominantly neutral,  are assumed to remain cooler and are not assigned artificially high temperatures. Thus, the definitions of ``neutral'' in these two sections serve different roles: one relates to computing realistic ionization fractions, and the other governs our thermal state assignment.}. Within neutral regions, a power-law
temperature density relationship is assumed, denoted as $T=T_0 \:
\Delta^{\gamma-1}$, where $T_0=6000$ K and $\gamma=1.6$. The selection of these
thermal parameters in neutral regions is driven by the adiabatic cooling
behavior post \HI and \HeI reionization at $z < 6$. Conversely, in regions
undergoing \HeII ionization, higher temperatures are expected. Here, the
power-law temperature density relationship is also applied, with thermal
parameter evolution drawn from observations by \cite{gaikwad2021}. 
The measured temperatures in \cite{gaikwad2021} represent an average over both
ionized and neutral regions. Assuming that ionized regions follow the measured
values while neutral regions are colder could introduce a small bias. However,
our tests on temperature fluctuations along sightlines indicate that the effect
is small \citep[see][for details]{gaikwad2021}. Moreover, our approach
considers an extreme case of temperature fluctuations, leading to more
conservative constraints on the mean free path and photo-ionization rates.
Accounting for these effects more accurately would require full radiative
transfer simulations, which are computationally expensive and would
significantly limit the parameter space exploration. Given the complexity of
\HeII reionization, our method provides a practical and conservative approach to
handling temperature fluctuations.

Three sets of thermal parameters are employed: (i) the default ($T_0,\gamma$),
(ii) a cold model ($T_0-\delta T_0, \gamma+\delta \gamma$), and (iii) a hot
model ($T_0+\delta T_0, \gamma - \delta \gamma$). These three parameter
combinations represent the maximum uncertainty in \LmfpHeII and \GHeIIavg in our analysis. For
instance, the cold (hot) model yields systematically higher (lower) values of
\LmfpHeII and \GHeIIavg. Importantly, these three models represent extreme
cases. A realistic representation of temperature fluctuations would likely
reside between these extremes.   Each of these thermal parameter combinations
results in the generation of 3731 models. Consequently, a total of $3731 \times
3 = 11193$ models are generated at a given redshift. For every model, the true
mean free path \LmfpHeII is calculated  and utilized for subsequent analysis. It
is noteworthy that the if  \HeII is fully ionized in a model at a given
redshift then we use the thermal parameters from the \cite{gaikwad2021}
measurements.

For each combination of \logLmfpHeII, \logGHeIIavg, and thermal parameters at
a given redshift, we extract fields along 60000 skewers. Utilizing these
skewers, approximately 24000 \HeII \lya forest spectra are generated, matching
the observed redshift path length (see \S\ref{sec:observation}). This set of \lya forest
spectra is further divided into 1000 individual mocks, with each mock
encompassing 24 sightlines. During parameter measurement, each mock sample is
treated with equal significance as the observational data.

Within each mock sample, we replicate 24 sightlines to mimic the observed 24
HST-COS sightlines. The properties of the observed \HeII \lya forest are drawn
from Table 1 in \cite{worseck2019}. Convolving the mock spectra with the
HST-COS line spread function, which varies with the temporal position of the
observations, is accounted for\footnote{\url{https://www.stsci.edu/hst/instrumentation/cos/performance/spectral-resolution}}.
To match the resolving power $(R)$ of the corresponding gratings (G130M and
G140L), the spectra are resampled on a wavelength array. Poisson noise is
added to  the spectra by employing photon counts at each wavelength
pixel. These counts are computed based on observed \SNR, exposure time ($t_{\rm
exp}$), flux at 1500 Å ($f_{1500}$), slope of the QSO continuum ($\alpha$), and
HST-COS instrument properties. 

%\subsection{Appendix:Realistic Poisson noise addition to spectraa}

\subsection{HeII \lya forest statistics}
The primary statistics used for the measurement of \LmfpHeII and
\GHeIIavg is the cumulative distribution function of \HeII effective optical
depth (\taueffHeII). This robust quantity can be derived from observations even
when individual \HeII \lya lines remain unresolved in current HST-COS
observations. Analogous to \taueffHI at $z>5$, previous studies by
\cite{worseck2019,gaikwad2019} demonstrated increased scatter in \taueffHeII
at $z>3$, an observation not explainable by a uniform UVB model. This increased
scatter has been attributed to fluctuations in ionizing radiation during
reionization. Within our formalism,
these fluctuations in the ionizing radiation field are captured by the two free
parameters \LmfpHeII and \GHeIIavg, making the \taueffHeII CDF a suitable
statistics for our measurements

The \taueffHeII is calculated using the
expression $\tau_{\rm eff,HeII} = -\ln \langle F_{\rm HeII, Ly \alpha}
\rangle$, where $\langle F_{\rm HeII, Ly \alpha} \rangle$ represents the mean
\HeII \lya flux along the sightline. Consistent with \cite{worseck2019}, we
compute \taueffHeII over an averaging window of $\Delta z_{\rm \tau}=0.04$. To
account for noise uncertainty, we derive the uncertainty in \taueffHeII from the
provided uncertainty in mean flux. Non-detections are identified within a
redshift bin if $\langle F_{\rm HeII, Ly \alpha} \rangle < 2 \langle
\sigma_{\rm noise} \rangle$, where $\langle \sigma_{\rm noise} \rangle$
signifies the binned noise in the redshift bin. In such cases, the effective
optical depth is calculated as $\tau_{\rm eff,HeII} = -\ln \: 2 \langle
\sigma_{\rm noise} \rangle$. Our non-detection criterion slightly deviates from
that employed by \cite{becker2015,bosman2018}. While we have incorporated
non-detection methods from existing literature, our analysis suggests that the
choice of the non-detection method has only a small impact on the constraints
of \LmfpHeII and \GHeIIavg.

In our  analysis of the observational data, the determination of \taueffHeII
(and its associated uncertainty) involves maximizing the Poisson likelihood
functions \citep{worseck2019}. This procedure is applied to spectra assuming
Poisson-limited conditions, whereby the probability of a specific
\taueffHeII value occurring along each sightline is computed. The
value of \taueffHeII is adjusted iteratively until the product of these
probabilities is maximized along the sightline. Despite the apparent
dissimilarity between our simulation-based \taueffHeII calculations and
observations, they are consistent with each other. This consistency arises due
to the existence of 1000 distinct realizations in our model for every observed
sightline. The principle of the central limit theorem assures the convergence of
the \taueffHeII cumulative distribution function (CDF) for such a large number
of realizations. To ensure convergence of the \taueffHeII CDF, we have verified
that the minimum required number of realizations are 80 whereas we are using
1000 realization for drawing inferences.

We employ the cumulative distribution function of \taueffHeII as the primary
statistics for the comparison between our models and observational data.  This
approach allows for a direct comparison between the \taueffHeII CDFs from our
models and the observed data. Non-parametric tests are well-suited for this
purpose, as they are straightforward to implement and are applicable to smaller
sample sizes. Given the limited number of clean QSO sightlines available for
\HeII \lya forest observations (as already discussed in \S
\ref{sec:observation}), the non-parametric tests accommodate the relatively
small sample size of the observed spectra ($\sim 24$). Note that these tests
do not rely on assumptions about the intrinsic \taueffHeII distribution within
the population.  Additionally, non-parametric tests exhibit greater resilience
to data outliers.  This robustness and their freedom from assumptions about
the intrinsic distribution provide a motivation for utilizing the \taueffHeII CDF to
quantify \LmfpHeII and \GHeIIavg.

%---------------
\InputFigCombine{{Parameter_Sensitivity_to_Spectra}.pdf}{155}%
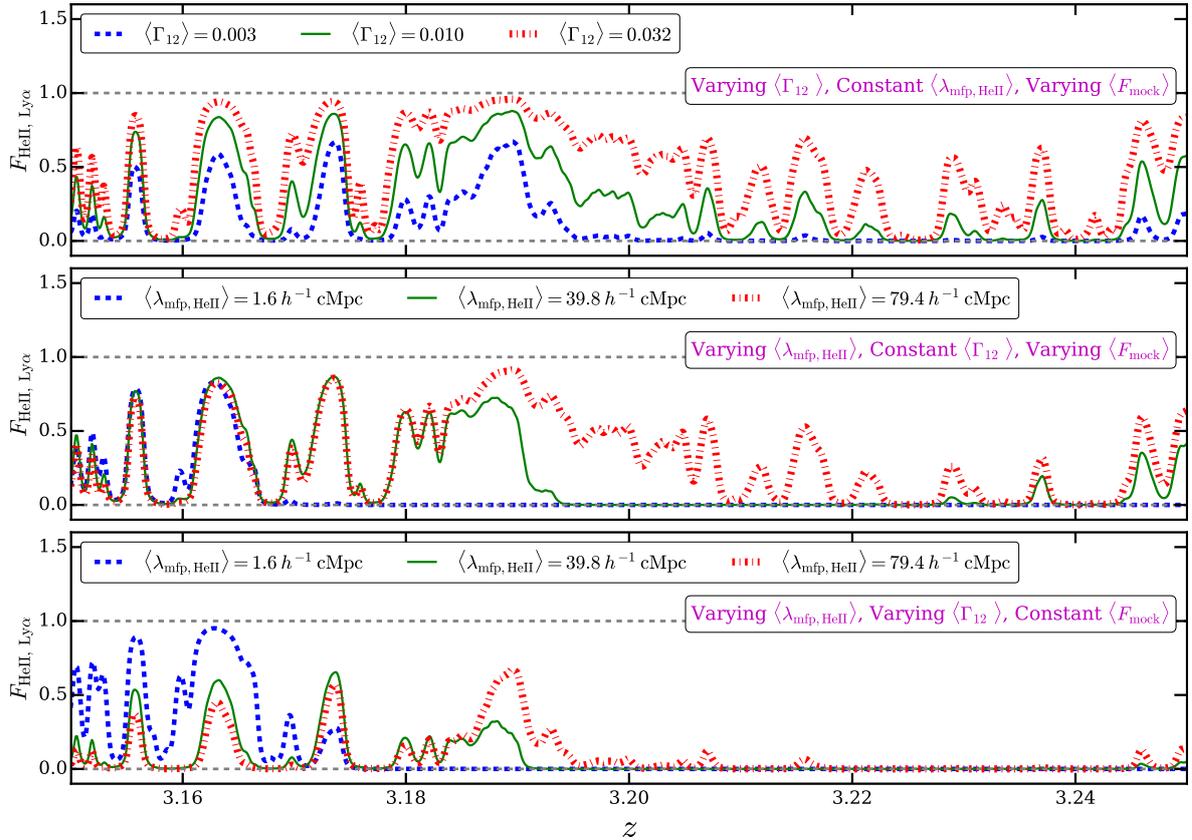
{ The top panel shows variation of HeII \lya flux with variation in average \HeII
    photo-ionization rate \GHeIIavg for constant average mean free path
    \LmfpHeII.  With the increase in  \GHeIIavg (red dashed curve), the flux
    increases as the neutral fraction decreases. However, the location where
    transmission spikes occur along a sightline does not change with \GHeIIavg.
    The middle panel shows variation of \HeII \lya flux with variation in average
    mean free path \LmfpHeII for constant \GHeIIavg.  With the increase in
    \LmfpHeII (red dashed curve), transmission spikes occur
    more frequently and at more locations along sightlines.  This
    is because a large mean free path allows photons to travel large distances
    that intersect with the line of sight at several locations.  For small mean
    free path (blue curve), the probability of ionizing the region along
    a sightline is small as photons can not travel large distances (see 
    \figref{fig:mfp-slice-comparison} for more details).  In the
    middle panel, even though \GHeIIavg is kept constant, the mean flux of the
    mock sample (\Fmeanmock) changes because of a large number of spikes in the  large mean
    free path model. For a fair comparison, one also needs to check if the
    transmission spikes occur when \GHeIIavg is changed such that mean flux of
    the mock spectrum is the same i.e., \Fmeanmock is constant. The bottom panel illustrates
    that even when \Fmeanmock is kept constant, the transmission spikes occur at more
    locations in large (red dotted curve) mean free path models than in the
    small mean free path models (blue and green curves). For the sake of visual
    clarity, we show simulated spectra excluding any observational effects such as
    LSF convolution and \SNR addition. In the rest of the paper all the
    results are shown for models that account for observational effects.
    \textit{Since changing \GHeIIavg changes the overall flux level, the
    median of \taueffHeII is sensitive to \GHeIIavg. On the other hand,
changing \LmfpHeII changes location and frequency of the occurrence of spikes. This
mainly affects the scatter in the \taueffHeII CDF (see \figref{fig:statistics-sensitivity}).}
}{\label{fig:spectra-sensitivity}}
%---------------
%---------------
\InputFigCombine{{Parameter_Sensitivity_to_Statistics}.pdf}{155}%
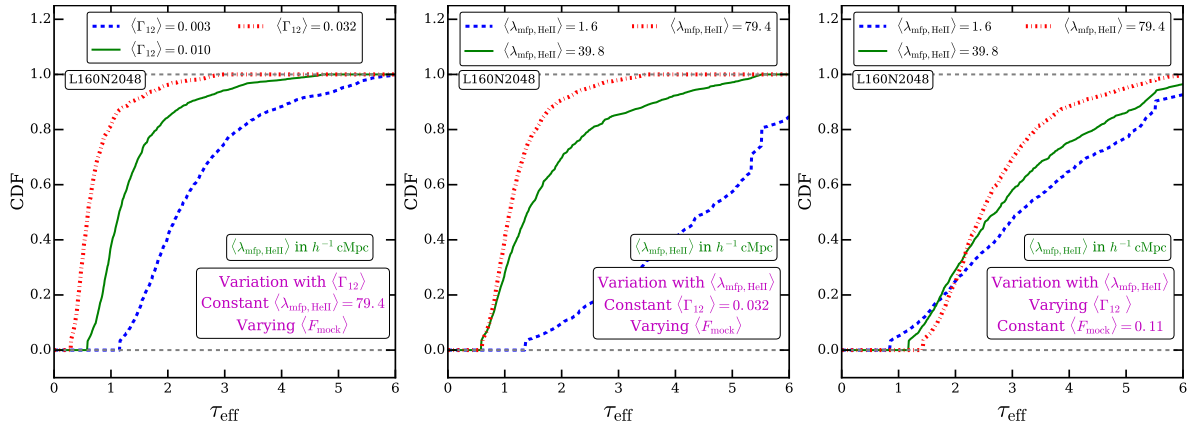
{ The left panel shows the sensitivity of the \taueffHeII CDF to variation in average \HeII
    photo-ionization rate \GHeIIavg for constant average mean free path
    \LmfpHeII.  With the increase in  \GHeIIavg, \taueffHeII decreases
    and the median of \taueffHeII  is systematically lower.  However, the
    shape of the \taueffHeII CDF (and thus the scatter in \taueffHeII) is not significantly
    affected by variation in \GHeIIavg. The middle panel shows the sensitivity of the
    \taueffHeII CDF to variation in average mean free path \LmfpHeII for
    constant \GHeIIavg.  With increase in \LmfpHeII, the scatter in
    \taueffHeII increases. However, the \taueffHeII CDF also shifts to the left
    because the mean flux \Fmeanmock is varying for this combination of parameters.
    The right panel shows the sensitivity of \taueffHeII CDF to variation in average
    mean free path \LmfpHeII for constant \Fmeanmock , where \Fmeanmock is
    the mean flux   of the entire mock sample. With increase in \LmfpHeII, the
    scatter in \taueffHeII increases and as a result, the shape of the \taueffHeII CDF
    changes. \textit{Thus, the median of the \taueffHeII distribution is sensitive to
        \GHeIIavg whereas the shape of the \taueffHeII CDF is sensitive to
    \LmfpHeII.} All the curves are shown for Sherwood simulation box
    L160N2048 at $z=3.2$.  The results are qualitatively similar for the \taueffHI
    CDF at $5 < z < 6$.
}{\label{fig:statistics-sensitivity}}
%---------------
We qualitatively examine the impact of varying \LmfpHeII and \GHeIIavg  on the
\HeII \lya flux along a single sightline in our model in
\figref{fig:spectra-sensitivity}. In the top panel, an increase in \GHeIIavg
leads to an overall elevation of flux levels due to decreased ionized
fractions. Notably, the locations of transmission spikes remain consistent
across varying \GHeIIavg values, as the mean free path \LmfpHeII remains
constant. In the middle panel of  \figref{fig:spectra-sensitivity}, we show  how
the \HeII \lya flux changes with variations in \LmfpHeII along the same
sightline. Larger \LmfpHeII values correspond to more frequent and dispersed
flux transmission spikes. This  is due to the increased likelihood of
encountering ionized regions along the sightline, due to the larger distances
traveled by ionizing photons in models with larger mean free paths. Consequently,
the morphology of reionization is significantly different, which is also evident
in \figref{fig:mfp-slice-comparison}. In the middle panel, the mean flux of the
mock spectra \Fmeanmock varies due to the increased appearance of spikes in
models with larger mean free paths, even with \GHeIIavg held constant. To ensure
a fair comparison, the bottom panel of \figref{fig:spectra-sensitivity}
demonstrates that even when maintaining a constant \Fmeanmock (through \GHeIIavg
variation), models with larger mean free paths exhibit spikes at numerous
additional locations.

We proceed to examine the impact of \GHeIIavg and \LmfpHeII on the cumulative
distribution function of \taueffHeII. We focus on the median and the tail of the
distribution, which together characterize its behavior well. The median value of
the distribution is sensitive to the neutral fraction of the IGM, while the
scatter of \taueffHeII  is sensitive to the  morphology of reionization.
Unsurprisingly, variations in the average photo-ionization rate \GHeIIavg affect
the neutral fraction and hence the median of the \taueffHeII distribution. The scatter in \taueffHeII is primarily due to three
contributions: (i) fluctuations within the cosmological density and velocity
fields, (ii) fluctuations in the photo-ionization rate, and (iii) temperature
fluctuations.  The influence of temperature fluctuations arises due to the
temperature dependence of the recombination rate coefficient. As mentioned
above, we assume a simplified two-zone model for the IGM temperature that
accounts for  temperature fluctuations in a somewhat  conservative way.  By varying
\GHeIIavg, \LmfpHeII and thermal parameters, our models encompass fluctuations
in the neutral fraction due to fluctuations in the cosmological density and
peculiar velocity fields, the photo-ionization rate fields and the temperature
fields.  We therefore expect variations in both the median and the scatter of the
\taueffHeII distribution across our models.

The left panel of \figref{fig:statistics-sensitivity} illustrates the influence
of varying \GHeIIavg on the \taueffHeII CDF while keeping \LmfpHeII constant. As
\GHeIIavg increases, the neutral fraction decreases and the mean flux increases,
leading to a decrease in \taueffHeII.  Consequently, the distribution
systematically shifts towards lower values of \taueffHeII, resulting in a
smaller median \taueffHeII. It is worth noting that the shape of the \taueffHeII
CDF that reflects the scatter in \taueffHeII, remains relatively similar. This can
be attributed to the fact that changing \GHeIIavg, while maintaining \LmfpHeII
fixed, impacts the overall flux level (as shown in
\figref{fig:spectra-sensitivity}), yet does not induce any variations in the
reionization morphology.  The middle panel of
\figref{fig:statistics-sensitivity} demonstrates the impact of varying \LmfpHeII
on the \taueffHeII CDF. For a fixed \GHeIIavg, larger \LmfpHeII values allow the
ionization of regions over larger distances.  Consequently, more regions along
sightlines exhibit transmission spikes, leading to an overall increase in
mean flux and a reduction in \taueffHeII for models with larger \LmfpHeII. This
shift in the median \taueffHeII towards lower values is a result of the enhanced
probability of encountering ionized regions along random sightlines in models
with larger \LmfpHeII, approaching a more homogeneous ionizing radiation field.
This, in turn, reduces the scatter in the \taueffHeII CDF. Conversely, in models
with smaller \LmfpHeII, random sightlines are less likely to pass through
ionized regions, which, in some cases, results in lower \taueffHeII, while
predominantly yielding higher \taueffHeII values when passing through neutral
regions. This leads to increased scatter in the \taueffHeII CDF.  The right
panel of \figref{fig:statistics-sensitivity} is similar  to the middle panel,
albeit with varied \GHeIIavg such that  the mean flux of the mock spectra
\Fmeanmock remains constant. The right panel underscores that \LmfpHeII
primarily affects the scatter in \taueffHeII, even in scenarios where \Fmeanmock
is held constant.

We emphasize that the commonly employed uniform UVB models found in the
literature are characterized by \LmfpHeII values significantly larger than the
simulation box size $L_{\rm box}$. Within these uniform UVB models, the scatter
in \taueffHeII primarily arises from fluctuations in the cosmic density and
velocity fields. Consequently, uniform UVB models exhibit  less scatter in the
\taueffHeII CDF when compared to fluctuating UVB models. Our fluctuating
models, on the other hand, exhibit a similar trend: as \LmfpHeII increases, the
scatter in the \taueffHeII CDF decreases. This behavior emerges because the
increasing \LmfpHeII values in our models lessen the impact of fluctuations in
the ionizing radiation field on the overall scatter within the \taueffHeII CDF.
In summary, the variation in \GHeIIavg influences the median of the \taueffHeII
CDF, while changes in \LmfpHeII modulate the scatter of this distribution.  In
subsequent analysis, these properties of the \taueffHeII CDF are effectively
used  to impose constraints on the mean free path and photo-ionization rate
from observational data.

%====================================================
\section{Results}
\label{sec:results}
%---------------
\InputFigCombine{{Parameter_Constraints_HeII_3_column}.pdf}{155}%
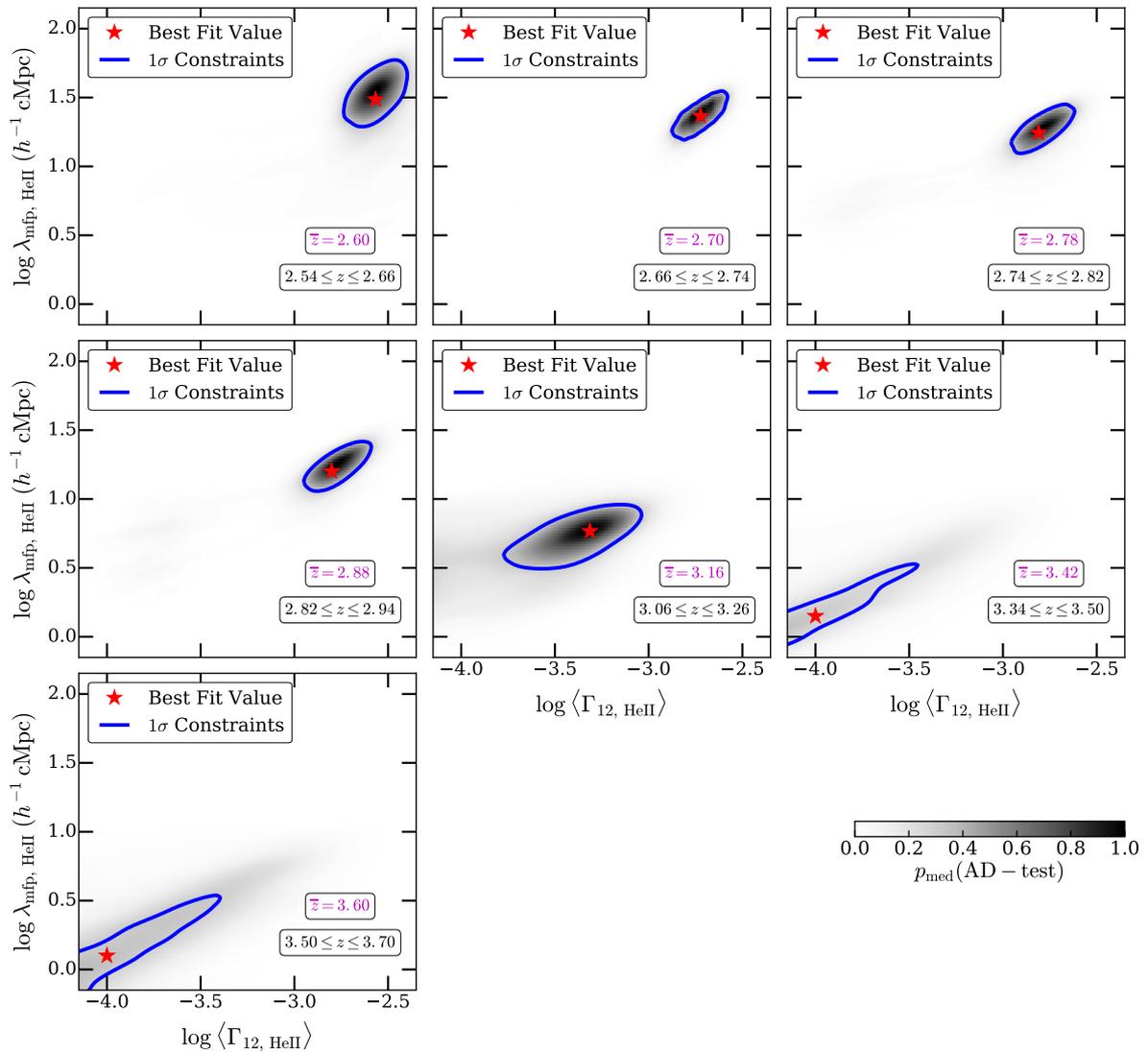
{Each panel shows the constraints on \LmfpHeII -\GHeIIavg in 7 different redshift
bins at \zrange{2.54}{3.70}. The constraints are obtained by comparing the
observed \taueffHeII CDF with that from the simulations using the non-parametric
Anderson-Darling test.  The red stars show the best fit values of
\LmfpHeII-\GHeIIavg while blue contours show the $1\sigma$ constraints on
\LmfpHeII-\GHeIIavg. The color scheme in each panel shows the median $p$ value
between observed and simulated \taueffHeII CDF.  The median $p$ value for each
combination of \LmfpHeII-\GHeIIavg is calculated from 1000 mock \taueffHeII
CDFs. The best fit values correspond to a model with maximum $p_{\rm med}$. The
$1\sigma$ contours correspond to $p_{\rm med}=0.32$ (see \S
\ref{subsec:parameter-constraints}) i.e., any model with $p_{\rm med}>0.32$ is
consistent with the data within $1\sigma$. The $1\sigma$ contours shown in
each panel also account for the thermal parameter uncertainties. The plot
clearly shows that \LmfpHeII and \GHeIIavg are evolving with redshift. Due
to the  number of non-detectections at $z>3.34$, we can only place
limits on \LmfpHeII and \GHeIIavg in the last two redshift
bins.}{\label{fig:parameter-constraints-multiple-redshift}}
%---------------
\InputFigCombine{{Best_Fit_CDF_HeII}.pdf}{155}%
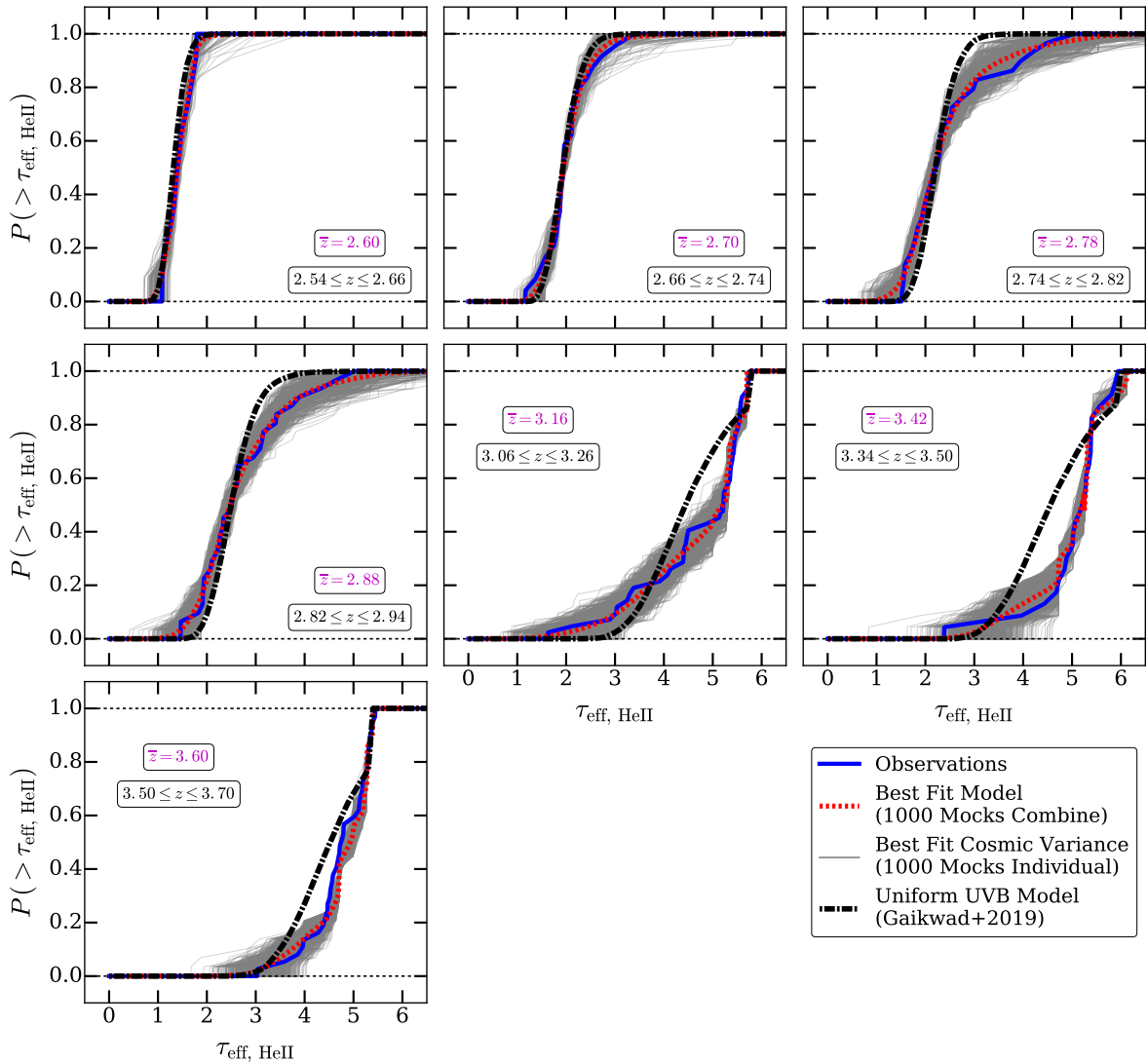
{ Each panel shows a comparison of \taueffHeII CDF between observations (blue curve)
    and the best fit model (red curve and gray curves). The \taueffHeII CDFs for uniform 
    UVB models are shown by black dashed curves \citep{gaikwad2019}.  The gray curves represent the
    best fit \taueffHeII CDF for each mock sample. Each mock sample has 
    the same redshift path length as the observations. 
    However, the skewers in each mock sample are different from each other.
    Thus, the gray curve represents the cosmic variance of the \taueffHeII CDF.  We vary
    \logLmfpHeII and \logGHeIIavg as free parameters and compute the AD test $p$
    value between simulated (the red curve that is obtained from 1000 mock samples) and
    observed \taueffHeII CDF. 
    The best fit model corresponds to the model with the maximum $p$ value. The uniform 
    UVB model matches the observations at $z<2.74$. However, it fails to reproduce the 
    \taueffHeII CDF at $z>2.74$. Our best fit model on the other hand reproduces the 
    median and scatter of the desired \taueffHeII distribution in all the redshift bins.
}{\label{fig:best-fit-model}}
%---------------

In order to measure \LmfpHeII and \GHeIIavg, we compare the \taueffHeII CDFs
derived from \excitecode simulations and those obtained from HST-COS
observations for seven distinct redshift bins. The methodology used here follows
the approach outlined in \cite{gaikwad2023} for their analysis of hydrogen
reionization. In this section,  we first briefly describe how we constrain
parameters and estimate the  associated uncertainty.  Following this, we explore
the implications of our measurements in the broader context of \HeII
reionization.

%========================
\subsection{Parameter Constraints}
\label{subsec:parameter-constraints}
We perform non-parametric Anderson-Darling (AD) statistical tests to compare the
observed \taueffHeII CDF with the simulations. For each model, we generate 1000
simulated mocks, ensuring they match the redshift path length and noise properties of the
observations. Utilizing AD statistics, we compute 1000 $p$ values that quantify
the similarity between the simulated and the observed \taueffHeII distribution. The
median $p$ value ($p_{\rm med}$) is then computed from these 1000 $p$ values for
each model. The optimal parameter values correspond to the model exhibiting the
highest $p_{\rm med}$. To establish the $1\sigma$ uncertainty, we require that
$p_{\rm med} \geq p_{\rm th}$, with $p_{\rm th}$ serving as the threshold. This
approach has been validated for parameter constraints in our previous work
\citep{gaikwad2023}, where we demonstrated the ability to recover parameters
within $1\sigma$.  Furthermore, in the same study, we validated $p_{\rm
th}=0.32$ as the appropriate choice with a bootstrap analysis involving 10000
samples from a self-consistent radiative transfer simulation.

In \figref{fig:parameter-constraints-multiple-redshift}, the constraints on
\logLmfpHeII and \logGHeIIavg are illustrated across seven distinct redshift bins
spanning \zrange{2.54}{3.70}. The gray color scheme represents the median $p$
value derived from the AD test between the observed and modeled \taueffHeII
CDF. Higher $p_{\rm med}$ values signify better agreement between  model
and data. Conversely, lower $p_{\rm med}$ values suggest inconsistency between
the simulated and the observed \taueffHeII CDF, implying distinct distributions. The
model yielding the highest $p_{\rm med}$ is identified as the best fit (red
star in each panel). Notably, as deviations from the best fit values increase,
$p_{\rm med}$ systematically decreases, reflecting the reduced agreement
between model and observations. The $1 \sigma$ statistical uncertainty of the
parameters is represented by blue contours, requiring $p_{\rm med} \geq 0.32$.

In \figref{fig:best-fit-model}, we present the comparison between the best-fit
and the observed \taueffHeII CDF. Gray curves represent the \taueffHeII CDF for
each of the 1000 mock samples generated for each model.  Additionally, we show a
uniform UVB model, depicted by the black curve, it displays the \taueffHeII CDF for a model 
without fluctuations in the ionizing radiation field \citep{gaikwad2019}.
Notably, the uniform UVB model diverges from the observed \taueffHeII CDF for $z
\geq 2.78$, particularly in regions with \taueffHeII $> 2.8$. This discrepancy
is expected due to the uniform UVB model's assumption of \LmfpHeII $\gg L_{\rm
box}$. In contrast, the gray curves representing the best-fit \excitecode model
match well within the  scatter in the observed \taueffHeII. This success can be
attributed to the varying mean free path, \LmfpHeII, in our model which is
consistently shorter than our simulation box size  ($ < 160 \: h^{-1} \:
{\rm cMpc}$). The higher density of gray curves near the observed \taueffHeII CDF
signifies increased consistency between the mock samples and observations. The
spread of gray curves across the \taueffHeII CDF encapsulates the line-of-sight variations due to the cosmic variance. To enhance clarity, we combine the \taueffHeII CDF
from 1000 mock samples into a red curve. This combined \taueffHeII CDF is used
solely for visualization purposes and does not contribute to the parameter estimation.
Notably, the red curve agrees well with the observed \taueffHeII CDF,
underscoring the ability of our models to reproduce both the median and scatter
observed in \taueffHeII.

For a realistic measurement of parameters, it is essential to consider the sources
of uncertainties inherent in our analysis. These uncertainties stem from three
key sources: (i) modeling uncertainty, (ii) cosmic variance, and (iii)
observational uncertainty. When modeling \HeII photo-ionization rate
fluctuations, our approach involves assuming default values for various
parameters such as thermal parameters ($T_0,\gamma$), halo mass cutoff ($M_{\rm
    cutoff}$), halo mass-emissivity power law index $(\beta)$, and mean free
path and photo-ionization rate dependence $(\zeta)$.  We have comprehensively
considered those  uncertainties in our final measurements.  Further details
regarding the impact of various uncertainties on the measured parameters can be
found in  appendix \ref{app:parameter-uncertainty}.
Here, we provide a brief summary of how different uncertainties affect our
measured parameters.

In \figref{fig:parameter-uncertainty-modeling}, we present an illustration of
how different parameters influence the constraints on \LmfpHeII and \GHeIIavg.
The parameter range explored to assess the impact on these quantities is
outlined in \tabref{tab:simulations}.  Among the sources of modeling uncertainty in our
analysis, the primary contributor is the uncertainty associated with thermal
parameters.  Conversely, uncertainties arising from other parameters, such as
$M_{\rm cutoff}$, $\beta$, and $\zeta$, have a relatively minor impact.  This
uncertainty can be attributed to the dependence of the recombination rate coefficient on the thermal parameters  that changes the neutral fraction
systematically.  Consequently, the uncertainty due to thermal parameters
results in systematic shifts of the $1\sigma$ contours.

\figref{fig:parameter-uncertainty-cosmic-variance} presents an analysis of the
impact of cosmic variance on the constraints of \LmfpHeII and \GHeIIavg. In
order to assess this influence, we use the 16$^{\rm th}$ and 84$^{\rm th}$
percentiles of the $p$ values. This approach enables the encapsulation of the
$1\sigma$ scatter evident across the 1000 mock samples. The figure further
demonstrates that this has only a marginal effect on the uncertainties
associated with the final measured parameters. Note that the
influence of cosmic variance becomes slightly more pronounced at higher
redshifts, a trend that can be attributed to the  reduced number of
sightlines, which in turn leads to increased sample variance.

\begin{table*}
\centering
\caption{The table shows the  measurements of the \HeII photo-ionization rate
    (\GHeIIavg, in $10^{-15} \; \mathrm{\ s}^{-1}$), mean free path (\LmfpHeII, in
    $h^{-1} \; \mathrm{cMpc}$), \HeII neutral fraction (\fHeII), emissivity at 228
    \text{\AA} ($\epsilon_{228}$, in $\mathrm{erg \: s^{-1} \: cMpc^{-3} \: Hz^{-1}}$) 
	and \HeII ionizing photon emission rate ($\dot{n}$, in $\mathrm{s^{-1} \: cMpc^{-3}}$) 
    with total $1\sigma$ uncertainty (i.e., including modeling,
    cosmic variance and observational uncertainties) \vspace{3mm}}
\resizebox{\textwidth}{!}{
\begin{tabular}{cccccc}
\hline  \hline
Redshift    & $\langle \Gamma_{\rm 15, HeII} \rangle$  & $\lambda_{\rm mfp, HeII}$ & $\langle f_{\rm HeII} \rangle$ & $\epsilon_{\rm 228}$ & $\dot{n}$ \\
\hline \noalign{\vskip 5pt} 
$2.60\:\pm\:0.06$  &  $2.723^{\: +1.446}_{\: -0.944}$  &  $33.884^{\: +26.372}_{\: -14.830}$  &  $8.749^{\: +17.408}_{\: -5.352} \: \times \: 10^{-3}$ &  $6.849^{\: +2.561}_{\: -2.031} \times 10^{23}$  &  $5.168^{\: +4.975}_{\: -2.371} \times 10^{49}$  \\ \vspace{2mm}
$2.70\:\pm\:0.04$  &  $1.862^{\: +0.829}_{\: -0.574}$  &  $23.442^{\: +12.866}_{\: -8.307}$   &  $1.456^{\: +2.356}_{\: -0.848} \: \times \: 10^{-2}$  &  $6.409^{\: +1.715}_{\: -1.522} \times 10^{23}$  &  $4.836^{\: +3.921}_{\: -2.000} \times 10^{49}$  \\ \vspace{2mm}
$2.78\:\pm\:0.04$  &  $1.622^{\: +0.833}_{\: -0.550}$  &  $18.621^{\: +10.219}_{\: -6.598}$   &  $2.198^{\: +4.411}_{\: -1.362} \: \times \: 10^{-2}$  &  $6.733^{\: +1.417}_{\: -1.357} \times 10^{23}$  &  $5.081^{\: +3.705}_{\: -1.960} \times 10^{49}$  \\ \vspace{2mm}
$2.88\:\pm\:0.06$  &  $1.718^{\: +0.912}_{\: -0.596}$  &  $17.179^{\: +9.736}_{\: -6.214}$    &  $2.263^{\: +5.512}_{\: -1.305} \: \times \: 10^{-2}$  &  $7.338^{\: +1.544}_{\: -1.479} \times 10^{23}$  &  $5.537^{\: +4.038}_{\: -2.136} \times 10^{49}$  \\ \vspace{2mm}
$3.16\:\pm\:0.10$  &  $0.398^{\: +0.557}_{\: -0.232}$  &  $5.248^{\: +4.302}_{\: -2.228}$     &  $5.964^{\: +1.207}_{\: -1.732} \: \times \: 10^{-1}$  &  $4.842^{\: +0.373}_{\: -0.693} \times 10^{23}$  &  $3.654^{\: +1.968}_{\: -1.245} \times 10^{49}$  \\ \vspace{2mm}
$3.42\:\pm\:0.08$  &  $ < 0.355$   					   &  $ < 2.089$   						  &  $ > 0.669$   										   &  $ < 9.602                      \times 10^{23}$  &  $ < 7.246   					  \times 10^{49}$  \\ \vspace{2mm}
$3.60\:\pm\:0.10$  &  $ < 0.417$   					   &  $ < 2.173$   						  &  $ > 0.674$   										   &  $ < 10.016                     \times 10^{23}$  &  $ < 7.558   					  \times 10^{49}$  \\ 
\hline \hline
\end{tabular}
}
\\
\label{tab:param-measurements}
\end{table*}

The uncertainty in the observed \taueffHeII introduces considerable uncertainty
in our measurements. This uncertainty is mainly due to how factors like fitting
the continuum, subtracting the sky background, and the limitations in counting
photons are handled. To understand their impact, we systematically adjusted the
\taueffHeII values by $\pm 1\sigma$ in the observations used in our models to make
predictions.  \figref{fig:parameter-uncertainty-observations} illustrates
how this observational uncertainty affects the derived values of \LmfpHeII and
\GHeIIavg. As expected, this uncertainty causes a noticeable shift in the
$1\sigma$ contours of parameter constraints. We have accounted for the
uncertainties stemming from these three sources and summarized them in the final
uncertainty estimates presented in \tabref{tab:param-measurements}. When
combining these uncertainties, we sum up the systematic uncertainties arising
from modeling and observational uncertainties, while cosmic variance
uncertainties are included in quadrature. This comprehensive approach should
result in robust and realistic  estimated uncertainty of the  measured
parameters.

In our model, specific combinations of \LmfpHeII and \GHeIIavg parameters
uniquely define the spatial distribution of neutral (\HeII) fractions. This
allows us to determine the spatially averaged \HeII fraction based on the
constraints on the  \LmfpHeII-\GHeIIavg parameters. Using the uncertainties provided in
\tabref{tab:param-measurements} for these parameters, we compute the \HeII
fractions within our simulation, leading to the constraints on the \HeII
fraction (\fHeII), as outlined in \tabref{tab:param-measurements}. Similar to
our approach for \LmfpHeII-\GHeIIavg, we account for uncertainties arising from
modeling and observations when evaluating the uncertainties in the \HeII
fraction. In the following section, we explore the evolution of these three
parameters and discuss its implications for \HeII reionization.

\subsection{Parameter evolution and its implications for \HeII reionization}
\label{subsec:parameter-evolution}
\InputFigCombine{{Parameter_Evolution}.pdf}{152}%
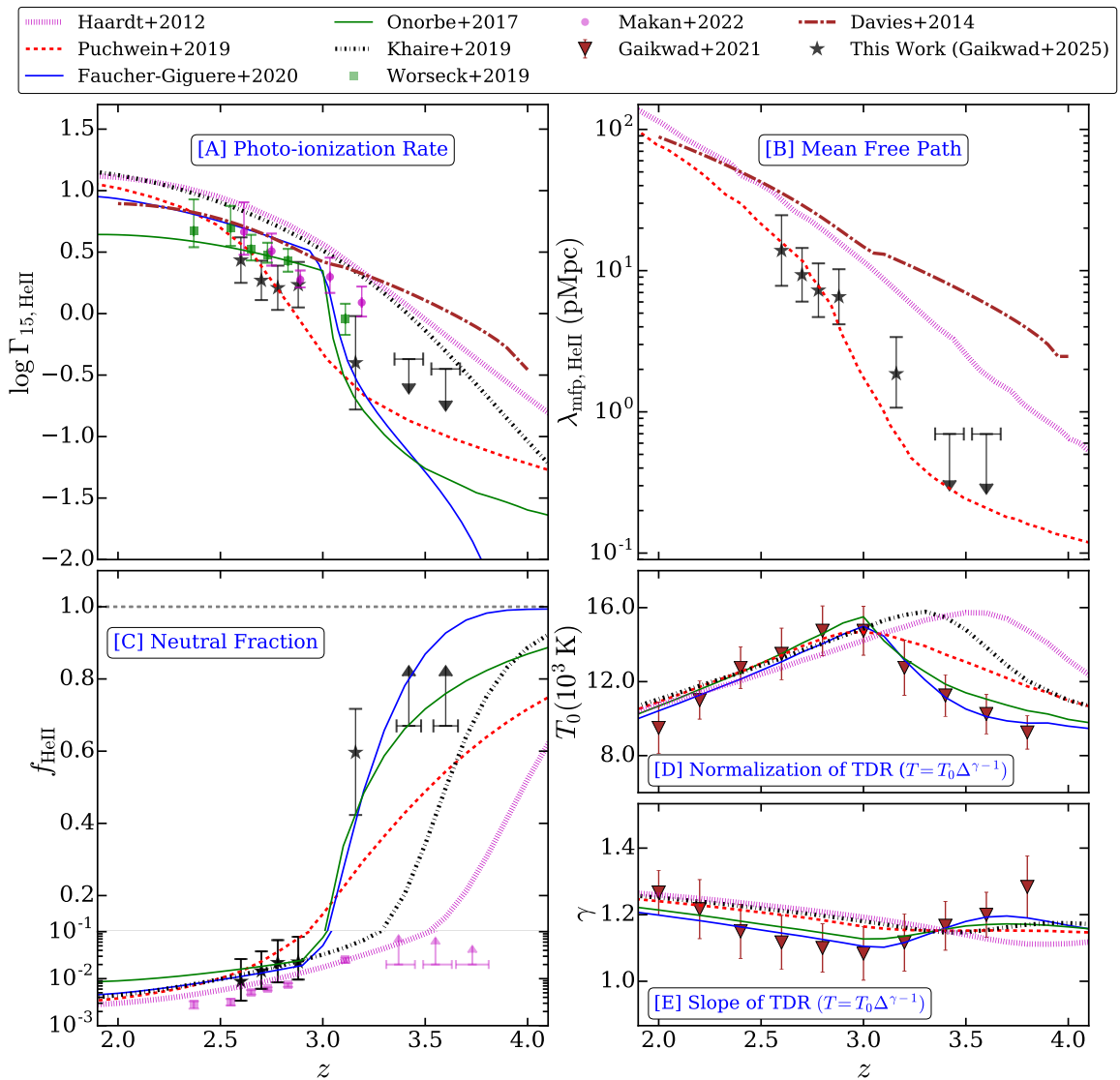
{ Panel A, B and C show the evolution of the \HeII photo-ionization rate
    (\GHeII, in units of $10^{-15} \: {\rm s^{-1}}$), mean free path of \HeII
    ionizing photons (\LmfpHeII) and \HeII fraction (\fHeII), respectively. In
    panel A, we also show \GHeII measurements of
    \cite{worseck2019,makan2022}. Contrary to our study, the \GHeII in these works has been measured
    assuming uniform UVB models (i.e., the mean free path is assumed to be much larger
    than simulation box size). As a result,  our best fit \GHeII
    measurements are systematically lower than that in the literature but they are
    still consistent within $1\sigma$. The uncertainties in our \GHeII
    measurements are usually larger because of the fluctuations in the UVB.  
    The different curves in  panel A show the \GHeII
    evolution for different UVB models available in the literature. The \GHeII
    evolution is consistent with the late \HeII reionization UVB models of
    \cite{davies2014,onorbe2017,puchwein2019,faucher2020}. The panel B shows the evolution
    of the mean free path measured in this work. The mean free path evolves rapidly
    between $z \sim 3.16$ and $z \sim 2.88$ indicating the ongoing process of
    \HeII reionization.  The mean free path evolution in \cite[][red
    dash curve]{puchwein2019} is in good agreement with our measurements
    (maximum differences $1.2 \sigma$). However, \cite{haardt2012,davies2014} systematically
    predict larger mean free path in their models. Panel C shows the inferred
    evolution of the \HeII fraction obtained in this work. The \fHeII
    constraints in \cite{worseck2019} are obtained assuming uniform UVB models
    hence the errorbars are small. Our \fHeII constraints are consistent with
    that of \cite{worseck2019} at $z<3$ while our best fit \fHeII are systematically
    higher at $z>3$. The \fHeII is consistent with the late and rapid \HeII
    reionization  models of \cite{onorbe2017,faucher2020}. The thermal parameter
    evolution from \cite{onorbe2017,faucher2020} is in good
    agreement with that from \cite{gaikwad2021} (panel D and E). Thus the \LmfpHeII, \GHeIIavg,
    \fHeII and thermal parameter evolution favors a scenario in which \HeII
    reionization is rather late and rapid. In all the UVB models, except \cite{davies2014}, we rescaled the
    photo-heating rates to match the IGM temperature evolution 
    \cite{gaikwad2021}.
}{\label{fig:parameter-evolution}}
%---------------
\InputFig{{Emissivity_ndot_Evolution_HeII}.pdf}{155}%
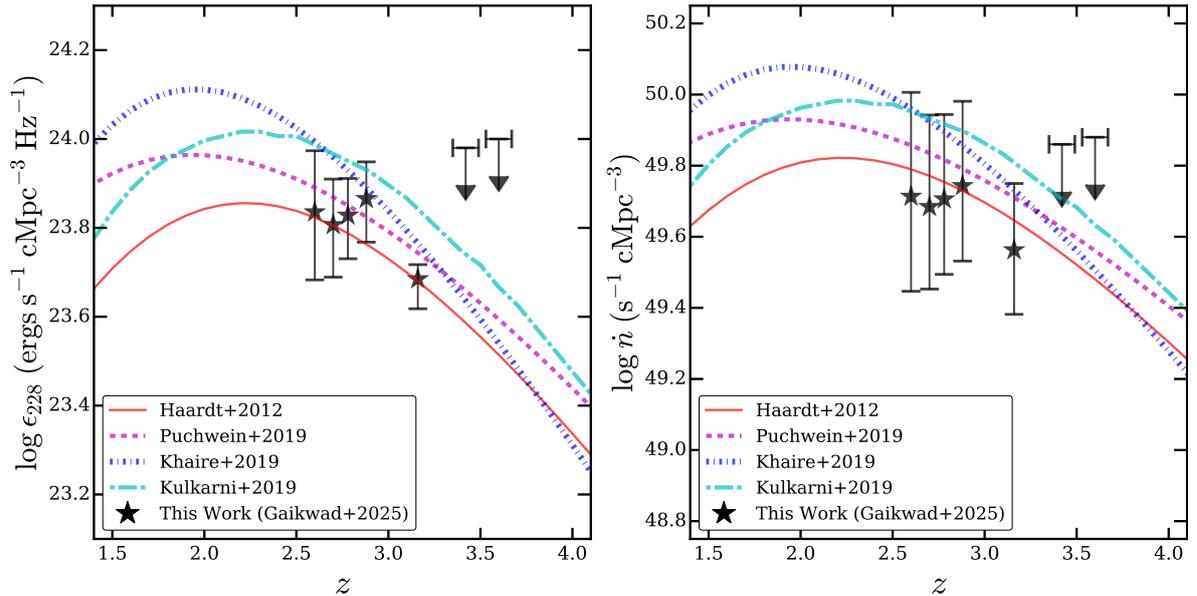
{The left panel shows evolution of emissivity at \HeII ionizing frequency 
($\epsilon_{\rm 228}$) obtained in this work. The emissivity is determined
using constraints on the mean free path and photo-ionization rate using 
absorption limited approximation.  The right panel shows the evolution of the
ionizing photon emission rate $\dot{n}$. The $\epsilon_{\rm 228}$ and $\dot{n}$
show relatively less evolution at all redshifts. The solid red, magenta dash 
and blue dotted curves shows the $\epsilon_{\rm 228}$ and $\dot{n}$ 
evolution from the UVB models of \cite{haardt2012,puchwein2019,khaire2019a}, respectively.
The cyan dot-dash curve is obtained by \cite{kulkarni2019b} using updated QSO 
luminosity functions. The emissivity and $\dot{n}$ evolution obtained in this 
work seems to be in reasonable agreement with that used in UVB models or obtained
from QSO luminosity functions.
Due to upper limits on  \LmfpHeII and 
\GHeIITW, we can only place upper limits on $\epsilon_{\rm 228}$ and $\dot{n}$.
Similar to Fig. \ref{fig:parameter-evolution}, the \HeII photo-heating rates in all UVB models are rescaled to match the IGM temperature evolution of \cite{gaikwad2021}.
}{\label{fig:emissivity-ndot-evolution}}
%---------------

\figref{fig:parameter-evolution} shows the evolution of the spatially
averaged photo-ionization rate (\GHeIIavg denoted by \GHeIITW, panel A), mean
free path (\LmfpHeII, panel B) and \HeII fraction (\fHeII, panel C).
Panel A shows the average photo-ionization rate (\GHeIITW), which remains fairly
constant at $z<3$ with only slight changes in redshift. A notable drop in the
best-fit \GHeIITW is observed between $3.06 \le z \le 3.26$. For $z>3.34$,
limitations in observations result in upper limits for \GHeIIavg, suggesting a
significant decrease at higher redshifts. We compare our findings with those of
\cite{worseck2019} and \cite{makan2022}. Notably, our \GHeIITW values tend to be
lower than those previously reported in the literature. This discrepancy could arise
due (i) differences in assumed thermal parameters, (ii) difference in simulation or
(iii) differences in not accounting for finite mean free paths. This
leads to smaller $\Gamma_{\rm HeII}$ values since $ \Gamma_{\rm HeII}  \propto
\lambda_{\rm mfp, HeII} $. Our \GHeIITW uncertainties are higher due to
fluctuations in \GHeII arising from the finite mean free path. Taking these
fluctuations of \LmfpHeII into account  contributes to the larger
uncertainties in \GHeIITW.  In  panel A of \figref{fig:parameter-evolution},
we also compare the \GHeIITW evolution with predictions from frequently used UVB
models. The early \HeII reionization models by \cite{haardt2012} and
\cite{khaire2019a} tend to predict higher \GHeIITW values at all redshifts,
contrasting our results. On the other hand, the late \HeII reionization UVB
models proposed by \cite{onorbe2017}, \cite{puchwein2019}, and \cite{faucher2020}
agree  better with our measurements. Finally, we also compare the \GHeIITW evolution
from \cite{davies2014} where a fluctuating mean free path model has been applied to \HeII 
reionization. The \GHeIITW predicted from \cite{davies2014} is found to be systematically
above our measured values.

The panel B of \figref{fig:parameter-evolution} shows the  redshift
evolution of \LmfpHeII.  Our measured \LmfpHeII values are generally smaller
than the size of our simulation box ($160 \: h^{-1} \: {\rm cMpc}$). In
the redshift range \zrange{2.66}{3.70}, the uniform model shown in
\figref{fig:best-fit-model} cannot reproduce the observed scatter in
\taueffHeII. On the other hand, our best-fit model with \excitecode matches the
observed \taueffHeII CDF quite well. At redshifts $z<3$, the evolution of
\LmfpHeII is relatively steady. However, at \zrange{3.06}{3.26}, \LmfpHeII
experiences a noticeable drop, accompanied by a decrease in \GHeIIavg due to the
ongoing \HeII reionization. This suggests that \HeII reionization is not
completed before  $z\sim 2.74$. For the redshift intervals \zrange{3.34}{3.50}
and \zrange{3.50}{3.70}, the influence of observational limitations, including
low signal-to-noise ratios and the small sample size, becomes significant. As a
result, we cannot constrain \LmfpHeII to smaller values during these periods.
This implies that the scatter in \taueffHeII for $z>3.34$ could be mostly due to
these observational limitations. Our upper limits on \LmfpHeII indicate a
significant change in its evolution from around $z \sim 3.2$ to $z \sim 3.8$.
The panel B of \figref{fig:parameter-evolution} also compares our
constraints with \LmfpHeII predictions from the UVB models by \cite{haardt2012} and
\cite{puchwein2019}. The early \HeII reionization model of \cite{haardt2012}
predicts consistently higher \LmfpHeII values, while the late \HeII reionization
model of \cite{puchwein2019} agrees better with our results.  Notably, the
changes in \GHeIIavg and \LmfpHeII go hand in hand at $z>3$, consistent with the
relationship $\lambda_{\rm mfp, HeII} \propto \Gamma_{\rm HeII}$ expected during
reionization. We should also emphasize that the fluctuating mean free path models
of \cite{davies2014} also predict evolution of the mean free path during \HeII reionization.
However, we find that their mean free path evolution is consistently above 
our measurements.

Panel C of \figref{fig:parameter-evolution} illustrates the evolution of
the \HeII fraction as derived from our study and from \cite{worseck2019}. Our
measured \fHeII exhibits a notable change between  $z \sim 2.88$ and $z \sim
3.16$. Our best-fit \fHeII, along with its $1\sigma$ uncertainty, tends to be
consistently larger than that reported in \cite{worseck2019}. This discrepancy
can be attributed to the consideration of ionizing radiation field fluctuations
in our analysis, which is absent in the approach by \cite{worseck2019}. Our
approach includes variations in the mean free path as well. As with \LmfpHeII
and \GHeIITW, lower limits on \fHeII can only be established in the last two
redshift bins. The different curves in  panel C of
\figref{fig:parameter-evolution} depict the \fHeII evolution predicted by five
different UVB models. Early \HeII reionization models such as those by
\cite{haardt2012} and \cite{khaire2019a} are inconsistent with our \fHeII
constraints with a significance of $2.9\sigma$ within the redshift range $3.06$ to
$3.26$. On the other hand, late \HeII reionization models like
\cite{onorbe2017} and \cite{faucher2020} are remarkably consistent with our
measured \fHeII evolution. While the \cite{puchwein2019} model also assumes
late \HeII reionization, it is more extended than the \cite{onorbe2017} and
\cite{faucher2020} models. Notably, the \cite{puchwein2019} model agrees with
our \fHeII evolution at $z<3$, but slightly deviates at $z>3$. In summary, the
evolution of \LmfpHeII, \GHeIITW, and \fHeII collectively indicate a scenario
of late and rapid \HeII reionization.  Our prior work \cite{gaikwad2021} found
results regarding the evolution of thermal parameters from \HI \lya
forest data consistent with the late \HeII reionization models presented by
\cite{onorbe2017} and \cite{faucher2020}.  
This has been shown in panel D and E of \figref{fig:parameter-evolution}. The findings of this work also
affirm the compatibility of our analysis with these independent earlier studies.
It is noteworthy that we have rescaled the photo-heating rates in the 
\cite{onorbe2017, puchwein2019, faucher2020} UVB models by a factor of 0.8, 0.9
and 0.7, respectively to match the evolution of thermal parameters
\citep[see][for details]{gaikwad2021}. This change in photo-heating rates mildly
affects the evolution of \GHeII, \LmfpHeII and \fHeII through the dependence of
recombination rates on temperature.

Given the measurements of the mean free path and photo-ionization rate, we are
in a position to translate these measurements into the constraints on the
emissivity and photon emission rate for \HeII. With the 'absorption
limited approximation' \citep{meiksin2005,gaikwad2023}, which is well-suited
for the later stages of \HeII reionization, we consider a scenario where the
mean free path of \HeII ionizing photons is shorter than the horizon size. This
approximation provides reasonable accuracy during the process of reionization.
With this approximation, the angle-averaged UVB intensity $J(\nu)$ is linked
to the mean free path ($\lambda(\nu)$) and the emissivity ($\epsilon(\nu)$)
through the equation $J(\nu) = \epsilon(\nu) \: \lambda(\nu) / 4\pi$.
We note that this assumption simplifies our calculation by 
treating the ionizing background as being sourced primarily by nearby emitters 
within one mean free path. This should be a reasonable approximation during HeII 
reionization and even shortly after, when the mean free path is still 
relatively short and the radiation field remains spatially inhomogeneous. 
Relaxing this approximation and including full radiative transfer would modify 
the relation between emissivity and the measured photo-ionization rate by 
accounting for the cumulative contribution of distant sources and redshifting 
of photons. However, this requires  self-consistent time dependent radiative
transfer which is beyond the scope of this work. Given the current level 
of uncertainty in our measurements and the dominant role of fluctuations 
in the ionizing background, we consider the local source approximation 
to be a reasonable choice for our analysis here.
The spatially averaged \HeII photo-ionization rate can be
obtained as,
\begin{equation}%
{\label{eq:absorption-limited-approximation}}
\langle \Gamma_{\rm HeII} \rangle = \int \limits_{\nu_L}^{\infty} \frac{\sigma(\nu) \; \lambda(\nu) \; \epsilon(\nu)}{h_p \nu} \;\; d\nu,
\end{equation}
where $h_p$ is the Planck constant, $\nu_L$ is \HeII ionization frequency (228 $\rm
\AA$) with $\sigma(\nu) = \sigma_{L}(\nu/\nu_L)^{-3}$ and $\sigma_L = 1.588
\times 10^{-18} \; {\rm cm}^2$ \cite{verner1994}. We assume that emissivity and
mean free path scale as power-laws of the form $\epsilon(\nu) = \epsilon_{228}
\: (\nu/\nu_L)^{-\alpha_{\rm s}}$ and $\lambda(\nu) = \lambda_{\rm mfp,HeII} \:
(\nu/\nu_L)^{3(\beta_{\rm HeII} -1)}$. We use $\alpha_{s} = 2.0 \pm 0.6$ and
$\beta_{\rm HeII} = 1.3 \pm 0.05$ consistent with the literature \citep{shull2012a,
lusso2014, stevans2014, lusso2015, tilton2016, lusso2018, gaikwad2019}.  The
emissivity at $228$ \text{\AA}  (in units of $ 10^{24} \; {\rm ergs \; s^{-1}
\; cMpc^{-3} \; Hz^{-1}}$) is calculated using the following relation,
\begin{equation}%
{\label{eq:em-228-expression}}
\epsilon_{228} = \bigg( \frac{0.2546}{\mathcal{I}_{\epsilon, \nu}} \bigg) \; \bigg( \frac{\langle \Gamma_{\rm HeII} \rangle}{10^{-15} \; {s^{-1}}} \bigg) \bigg( \frac{10 \: {\rm pMpc}}{\lambda_{\rm mfp,HeII}}\bigg) \bigg( \frac{4}{1+z}\bigg)^{3}. 
\end{equation}
where $I_{\epsilon,\nu} = \alpha_{\rm s} - 3 \beta_{\rm HeII} +6$ and \GHeIIavg and \LmfpHeII are photo-ionization rate and mean free path constrained from the observations. We calculate the \HeII ionizing photon emission rate at a given redshift using, 
\begin{equation}%
{\label{eq:ndot-expression}}
    \dot{n} = \int \limits_{\nu_{\rm L}}^{\infty} \: \frac{\epsilon(\nu)}{h_{\rm p} \nu} \: d\nu = \frac{\epsilon_{\rm 228}}{h_{\rm p} \: \alpha_s} .
\end{equation}

The constraints on emissivity and ionizing photon emission rate obtained from
our formalism are depicted in \figref{fig:emissivity-ndot-evolution}. The
emissivity remains relatively constant within the redshift range
\zrange{2.54}{2.94}, as illustrated in the top panel, but exhibits a decline
around $z\sim 3.16$. This decline corresponds to the ongoing \HeII
reionization, where both \LmfpHeII and \GHeII are still evolving. Similarly,
the best fit values of the ionizing photon emission rate ($\dot{n}$) show
redshift evolution, as shown in the bottom panel. However, the uncertainty
associated with $\dot{n}$ is slightly higher due to uncertainties in
$\alpha_{\rm s}$. Within this uncertainty, $\dot{n}$ appears to remain
relatively consistent across redshifts \zrange{2.54}{3.16}. Additionally, we
compare the evolution of emissivity and photon emission rate obtained from
uniform UVB models in \figref{fig:emissivity-ndot-evolution}. Despite the
larger uncertainties, all uniform UVB models are broadly consistent with our
constraints. Notably, the models proposed by \cite{puchwein2019} exhibit
better agreement with our constraints compared to the other models. In the 
\cite{haardt2012,puchwein2019} UVB models employing one-zone radiative transfer, \HeII reionization ends at 
$z\sim2.8$.  It is
noteworthy that these uniform UVB models derive the emissivity evolution from
observed luminosity functions and spectral energy distributions of galaxies and
QSOs. Hence, the consistency of our constraints with the evolution of these
models suggests agreement with observational properties inferred from galaxy and QSO
surveys \citep{hopkins2007, palanque2013, reed2015, giallongo2015,
kulkarni2019b, shen2020, pan2022}.

Finally, we emphasize that uniform UVB models can not reproduce the observed
scatter in \taueffHeII at $z>2.74$, indicating the necessity of modeling
inhomogeneous \HeII reionization with a finite mean free path to explain the
observations. This suggests that \HeII reionization is still ongoing at $z>2.74$ .
This rather late completion of \HeII reionization has implications for using \HI
\lya as a cosmological probe, as temperature fluctuations persist in the IGM
long after reionization. Ongoing and forthcoming \HI \lya forest surveys such as
{\sc desi} and {\sc weave}, aimed at measuring the baryon acoustic oscillation
signal at $z<2.5$, should therefore account for large-scale temperature and UVB
fluctuations arising from \HeII reionization to accurately determine
cosmological parameters like the expansion of the universe.

\section{Discussion}
\label{sec:discussion}
In this section we discuss the limitations of our current study and scope for
future improvements. Firstly, our approach assumes that QSOs emit radiation
continuously throughout their lifetime and in all directions isotropically.
However, in reality, QSOs have a duty cycle and emit radiation only during
specific periods \citep{croft2004, khrykin2016,khrykin2017, khrykin2019,
khrykin2021}. Quantifying this duty cycle remains challenging.  The
measurements of the QSO luminosity function typically capture only active QSOs
at a given epoch, complicating efforts to determine the duty cycle accurately.
Moreover, QSOs often emit radiation in an anisotropic manner, typically in a
bipolar fashion, which may result in a somewhat heterogeneous morphology
\citep{moller1992, schirber2004, kirkman2008, furlanetto2011}. These effects
have not been considered in our analysis.

We now argue that these effects may exert a slightly smaller influence on \HeII
reionization. This is primarily because \HeII reionization is a global
phenomenon that affects the distribution of \HeII on intergalactic scales. What
matters most at any given point in the IGM is the distribution of reionization
sources in its vicinity.  Given that QSOs exhibit clustering, even if some QSOs
are inactive for periods, there remains a likelihood that other QSOs in the same
regions will be active, thereby driving  the reionization process
\citep{porciani2004, shen2009, white2012, eftekharzadeh2015, oogi2016,
laurent2017, rogriguez2017, timlin2018, greiner2021}. The main effect of not
including a duty cycle is that all sources are continuously active, effectively
increasing the number of contributing sources while reducing their individual
impact. In contrast, incorporating a duty cycle would result in fewer active
QSOs at a given time, each contributing more ionizing photons. This could
influence the shape of the QSO luminosity function \citep{basu2024}. Accurately
modeling the duty cycle requires tracking the time evolution of reionization
through self-consistent radiative transfer simulations. However, such
simulations would limit our ability to explore a broad parameter space using a
static reionization field, which is the main focus of this work. Alternatively,
one can include a duty cycle keeping the source density and QSO luminosity
function fixed but varying the parent population of halos. However this approach
requires to accurately account for clustering of QSO which would be beyond the
scope of current work. In future studies, we plan to investigate the effects of
the QSO duty cycle on \HeII reionization in greater detail.

A similar rationale applies to the anisotropic emission from QSOs. Since the
emission direction of QSOs is inherently random, on large scales, the ionized
bubbles may exhibit isotropic characteristics due to the clustering of QSOs
emitting photons in all directions. Additionally, while QSO emission may appear
anisotropic on small scales initially, continued reionization may ultimately
render them isotropic. This is because the \HeII atoms ionized closest to the
QSOs, may subsequently recombine. If this recombination occurs directly to the
\HeII ground state, it will emit \HeII ionizing photons in random directions,
facilitating the isotropic diffusion of ionizing photons.

In this study, our box size is limited to $160 h^{-1} \: {\rm cMpc}$. It is well
known that smaller box sizes may under-sample the highest density peaks,
particularly those associated with the most massive halos. Achieving convergence
in the properties of the \HeII \lya forest typically demands higher resolution,
necessitating computationally expensive simulations with a large dynamic range.
While the limited box size may influence parameter inferences to some extent, we
anticipate its impact to be moderate. This is because uncertainties in our
measured parameters are primarily driven by observed uncertainties in thermal
parameters and observational systematics, such as the number of sightlines and
the signal-to-noise ratio of the spectra. Thus, the clustering of 
QSOs in dense environments suggests that the impact of QSO duty cycle and chosen
anisotropic emission may be relatively minor in \HeII reionization. 

We emphasize  that non-equilibrium effects may become important, particularly at $z > 3$ where the photo-ionization timescale approaches the Hubble time. To assess the potential impact of time-delay effects due to non-equilibrium ionization, we performed an additional test in which the spatially varying $\Gamma_{\rm HeII}$ field within ionized regions was replaced by its mean value $\langle \Gamma_{\rm HeII} \rangle$, while keeping the underlying density and ionization fields fixed. We repeated this exercise for several representative values of $\lambda_{\rm mfp,HeII}$ from Table 2. We find that the cumulative distribution of $\tau_{\rm eff,HeII}$ changes very little under this modification, indicating that this statistic is primarily sensitive to the mean transmitted flux rather than the detailed spatial structure of $\Gamma_{\rm HeII}$. This suggests that, for $\tau_{\rm eff,HeII}$, large-scale background fluctuations are largely averaged out along the line of sight. However, quantities that depend more strongly on the small-scale structure of the transmission, such as the distribution of transmission spikes, are expected to be more sensitive to these fluctuations. A detailed analysis of such statistics would therefore be a valuable next step for further constraining the role of spatial variations in the ionizing background.

Finally,  We note that a fully self-consistent treatment accounting for non equilibrium effects would be more physically complete. However our choice is motivated by two practical and methodological considerations. First, radiative-hydrodynamic simulations that capture time-dependent ionization evolution are computationally expensive and would significantly limit our ability to explore a wide range of parameters, especially given the large variation in neutral fraction that needs to be sampled to match the data. Second, the true reionization history remains uncertain. By analyzing each redshift bin independently, we avoid imposing a specific reionization model and allow the data to guide the inference without strong priors on the temporal evolution. This is particularly important for a first measurement of the mean free path, where robustness and minimal modeling assumptions are preferred. In future, we aim to perform high dynamic range cosmological
radiative hydrodynamics simulations that incorporate these effects to further
investigate \HeII reionization.  However, based on our prior experience with \HI
reionization, performing such simulations would be exceedingly costly, likely
limited to only a few. The parameter estimation undertaken in this study will
greatly aid in calibrating these simulations, thus facilitating future research
efforts in \HeII reionization. This underscores the significance of our work for
advancing understanding of the second major phase transition in the universe.

%%====================================================
\section{Summary}
\label{sec:summary}
\HeII reionization is an important milestone in the history of the universe,
that is more accessible to observation than \HI reionization. Recent
advancements in observing the \HeII \lya forest highlight the significance of
patchy \HeII reionization effects. Characterized by the mean free path of \HeII
ionizing photons and the spatially averaged \HeII photo-ionization rate, the
patchiness of reionization poses a modeling challenge. In this study, we measure
the mean free path (\LmfpHeII), spatially averaged \HeII  photo-ionization rate
(\GHeIIavg), and \HeII fraction (\fHeII) in the redshift range
\zrange{2.54}{3.70} by comparing observed effective optical depth distributions
of \HeII with large number of patchy \HeII reionization models, varying these
parameters. Our analysis provides the first observational constraints on the
mean free path of \HeII ionizing photons and considers the uncertainty in \GHeII
due to patchiness of \HeII reionization. We find that the evolution of
\LmfpHeII, \GHeIIavg, and \fHeII agrees with uniform models by
\cite{onorbe2017,faucher2020}, suggesting a scenario of late and rapid \HeII
reionization. Consistency with previous analyses of thermal parameters from \HI
\lya forest lends further support to this unified and consistent picture. Below,
we provide a detailed summary of our findings.

\begin{itemize} 
\item  We have used \HeII effective optical depths derived from observations of the
    \HeII \lya forest with the Cosmic Origins Spectrograph on the Hubble Space
        Telescope, as presented in \cite{worseck2019}. Employing our newly
        developed code, \excitecode, we model patchy \HeII reionization,
        allowing efficient variation of \LmfpHeII and \GHeIIavg.  In
        \excitecode, we introduce fluctuations in \GHeII by iteratively varying
        the \GHeII field using a physically motivated expression for the local
        mean free path, following methods validated in previous studies
        \citep{davies2016,gaikwad2023}. In this work, we generate $\sim 15000$ patchy
        \HeII reionization models, varying \LmfpHeII, \GHeIIavg, and other
        modeling parameters at four different redshifts by post-processing
        outputs from the Sherwood simulation suite.

\item  We have simulated the \HeII \lya forest for various \LmfpHeII-\GHeIIavg
    parameter variations and have analyzed its properties. Our findings indicate that
        changes in \GHeIIavg influence the mean transmission level, while
        variations in \LmfpHeII impact the location of transmission spikes,
        altering reionization morphology. Moreover, we demonstrate that
        variation in \LmfpHeII affects the scatter in the \taueffHeII cumulative
        distribution function (CDF), while \GHeIIavg systematically shifts the
        \taueffHeII CDF without altering its scatter. Leveraging these
        characteristics of the \taueffHeII CDF, we constrain \LmfpHeII and
        \GHeIIavg using observed data.

\item We use a non-parametric Anderson-Darling test to simultaneously
    constrain \LmfpHeII and \GHeIIavg by comparing the observed \taueffHeII CDF
        with simulations.  The best-fit model is identified by maximizing the
        AD test $p$ value against the observed \taueffHeII. Confidence
        intervals at $1\sigma$ are determined using a predefined cutoff of
        $p>0.32$, established through random realizations and MCMC parameter
        recovery. Accounting for observational and modeling uncertainties, we
        particularly consider thermal parameter uncertainties, adopting
        conservative estimates derived from robust constraints by
        \cite{gaikwad2021}. The limited number of sightlines together with low SNR
        spectra at $z>3.34$ restrict our ability to constrain \LmfpHeII and
        \GHeIIavg, leading to upper limits.  Each \LmfpHeII-\GHeIIavg
        combination corresponds to a unique globally averaged \HeII fraction
        (\fHeII), allowing us to transform constraints into \fHeII, thereby
        characterizing the reionization history.

\item Our analysis shows a redshift-dependent evolution of the measured
    parameters within \zrange{2.54}{3.26}. While \GHeIIavg remains relatively
        constant from \zrange{2.54}{2.88}, our measurements are systematically
        lower than those in previous literature, attributed to differences in
        assumed thermal parameters. However, our increased uncertainty on
        \GHeII is due to marginalization over \LmfpHeII, capturing patchy \HeII
        reionization. For the first time, we have constrained the mean free path of \HeII
        ionizing photons (\LmfpHeII). Our measurements shows a systematic
        decrease in \LmfpHeII from $z=2.54$ to $z=3.70$, with a significant
        drop observed at $z>3.06$. This decreasing trend in \GHeII and \LmfpHeII 
        suggests an evolution in the size of ionized bubbles with redshift, 
        indicating an incomplete process of \HeII reionization.

\item In our analysis we see a significant evolution in \fHeII, similar to
    \GHeIIavg and \LmfpHeII. We consistently find higher values of \fHeII
    compared to the constraints by \cite{worseck2019}, that can be attributed to our
    consideration of patchy reionization rather than a uniform UVB. Previous
    estimates based on uniform UVB models often underestimated \fHeII due to
    their reliance on matching the mean flux rather than the effective optical depth
    distribution. Our results indicate \fHeII $\sim 0.6$ at
    \zrange{3.06}{3.26}, suggesting ongoing reionization at these redshifts.
    Despite observational limitations such as a limited number of sightlines and lower
    SNR, our study establishes upper limits on \GHeIIavg and \LmfpHeII, and
    lower limits on \fHeII beyond $z>3.26$, representing significant
    improvements over the existing literature and indicating notable evolution in
    reionization progression at these redshifts.
	We examine the evolution of \GHeIIavg, \LmfpHeII, and \fHeII by comparing
    them with uniform UVB models in the literature, including those by
    \cite{onorbe2017}, \cite{puchwein2019}, and \cite{faucher2020}. These
    models, where \HeII reionization is late and rapid agree well with our
    measured parameters. 
\end{itemize} 

The measured parameters in our study hold several significant
implications.  Consistency between our findings and previous work on thermal
parameter measurements based on \HI forest observations  supports the
consistent picture of late \HeII reionization that is completed by $z \sim
2.74$.  Our measured parameters provide valuable inputs for calibrating
cosmological radiative hydrodynamics simulations in future investigations.
Additionally, the delayed completion of \HeII reionization has implications for
Baryon Acoustic Oscillations (BAO) measurements. The temperature fluctuations
induced by the reionization process persist in the IGM long after reionization
concludes.  These effects are crucial considerations for deriving cosmological
parameters from BAO measurements in ongoing and upcoming surveys such as {\sc desi}
and {\sc weave}.

%---------------

%====================================================

%====================================================
%\vspace{-8mm}
\section*{Acknowledgments}
We thank the anonymous reviewers for their helpful suggestions and constructive feedback, which significantly improved the manuscript.
P.G. acknowledges funding from  IIT-Indore, through a Young Faculty Research Seed 
Grant (project: `INSIGHT'; IITI/YFRSG/2024-25/Phase-VII/02).
The \sherwood simulations and its post-processing were performed using the
Curie supercomputer at the Tre Grand Centre de Calcul (TGCC), and the DiRAC
Data Analytic system at the University of Cambridge, operated by the University
of Cambridge High Performance Computing Service on behalf of the STFC DiRAC HPC
Facility (www.dirac.ac.uk). This equipment was funded by BIS National
E-infrastructure capital grant (ST/K001590/1), STFC capital grants ST/H008861/1
and ST/H00887X/1, and STFC DiRAC Operations grant ST/K00333X/1. DiRAC is part
of the National E- Infrastructure.  Computations in this work were also
performed using the CALX machines at IoA.  Support by ERC Advanced Grant 320596
`The Emergence of Structure During the Epoch of reionization' is gratefully
acknowledged. MGH acknowledge the support of the UK Science and Technology
Facilities Council (STFC) and  the National Science Foundation under Grant No. NSF PHY-1748958.  Part of the work has been performed as part of the DAE-STFC collaboration ‘Building Indo-UK collaborations towards the Square Kilometre Array’ (STFC grant reference ST/Y004191/1).

\section*{Data Availability}
The data generated during this work will be made available upon reasonable request to the corresponding author.
%
%
%
%
%%\vspace{-6mm}
%%%%%%%%%%%%%%%%%%%%% REFERENCES %%%%%%%%%%%%%%%%%%
%
%% The best way to enter references is to use BibTeX:
\bibliographystyle{JHEP}
\bibliography{HeII_excite} % if your bibtex file is called example.bib
%
%%%%%%%%%%%%%%%%%%%%% APPENDIX %%%%%%%%%%%%%%%%%%
\appendix

\section{Convergence Tests}
\label{app:convergence-test}
To effectively model patchy \HeII reionization, large dynamic range simulation
boxes are essential, as the sources responsible, such as QSOs, are typically
found in massive halos. Achieving sufficiently large halo masses requires
large box sizes. However, the \HeII \lya forest is observed at a higher
resolution, typically 60 km/s. Additionally, given that the helium atom is four times
heavier than the hydrogen atom, typical Doppler broadening due to the gas temperature is
halved. While the current \HeII \lya forest resolution is not adequate for studying
the thermal state of gas from the \HeII \lya forest, it is crucial to assess
simulation convergence in both mass resolution and box size. Moreover, to
explore a broad parameter space, we typically model \GHeII fluctuations on
$N_{\rm Grid, \Gamma_{\rm HeII}} = 512^3$ grids, highlighting the importance of
confirming the adequacy of this grid size for achieving convergent results.

The left panels of \figref{fig:convergence-test-1} and
\figref{fig:convergence-test-2} illustrate the impact of varying box size on
the cumulative distribution function  of \taueffHeII at fixed mass
resolution. For the smallest box size ($L = 40 \: h^{-1} \: {\rm cMpc}$), there
is slightly larger scatter in \taueffHeII compared to larger boxes, indicating
the absence of some very massive halos. However, convergence in the \taueffHeII
CDF is observed with increasing box size. Notably, our Sherwood simulation
suite has a maximum box size of $160 \: h^{-1} \: {\rm cMpc}$, but larger
boxes without compromising
resolution may further improve \HeII reionization studies . Such simulations are planned for future work, given their
computational expense, which exceeds the scope of the current study.

In the right panels of \figref{fig:convergence-test-1} and
\figref{fig:convergence-test-2}, we explore the impact of varying mass
resolution while maintaining the same box size. The lowest resolution model,
L160N512, exhibits clear lack of convergence, while increasing the mass
resolution to L160N1024 and L160N2048 results in convergence of the \taueffHeII
CDF. For our parameter inference, we mainly use the L160N2048 models, which
offer the widest dynamic range available in the Sherwood suite.

In \figref{fig:excite-maps-resolution}, we display a simulation box slice
illustrating the impact of varying $N_{\rm Grid, \Gamma_{\rm HeII}}$ on \GHeII
fluctuations. All models in the figure  have the same box size and mass resolution.
For $N_{\rm Grid, \Gamma_{\rm HeII}}=64^3$, large \GHeII fluctuations are
accurately represented, but small-scale features exhibit artificial smoothing
due to the coarse resolution in density and hence local mean free path computation.
Increasing $N_{\rm Grid, \Gamma_{\rm HeII}}$ to $1024^3$ enhances small-scale
fluctuations, with convergence observed when $N_{\rm Grid, \Gamma_{\rm HeII}} >
256^3$. We adopt $N_{\rm Grid, \Gamma_{\rm HeII}} = 512^3$ throughout this work
to generate a sufficiently large number of models. A quantitative comparison of
\taueffHeII CDFs in these models at $z \sim 3.16$ and $z \sim 2.70$ is
presented in the right panels of \figref{fig:convergence-test-1} and
\figref{fig:convergence-test-2}, respectively, indicating well-converged
results for $N_{\rm Grid, \Gamma_{\rm HeII}} = 512^3$.
%---------------

\InputFigCombine{{Simulation_Resolution_Study_z_3.16}.pdf}{155}%
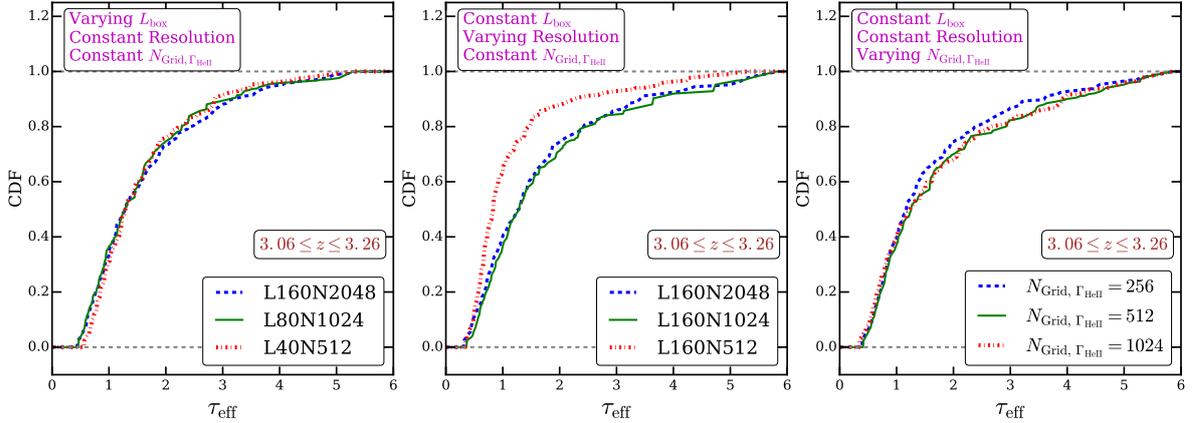
{ The left panel displays the variation of the \taueffHeII CDF with changes in
box size, for fixed mass resolution and \GHeII / \GHeIIavg map resolution.
Convergence of the \taueffHeII CDF is evident ($p_{\rm med} \sim 0.96$) for a
box size of $L = 160 h^{-1} \: {\rm cMpc}$. Notably, a fixed mean free path of
HeII ionizing photons ($\lambda_{\rm mfp, HeII}$) $= 31 h^{-1} \: {\rm cMpc}$ is
chosen for all models to ensure a fair comparison. The value of the mean free path is
chosen to be smaller than the smallest box size ($40 \: h^{-1} \: {\rm cMpc}$).
In the middle panel, convergence of the \taueffHeII CDF is shown for the
mass resolution of the simulation box, keeping  box size and
\GHeII / \GHeIIavg map resolution fixed. The \taueffHeII tends to be under-predicted
($p_{\rm med} \sim 0.03$) for the lowest mass resolution (L160N512) due to
inadequate density field convergence. However, the \taueffHeII CDF converges
relatively well for our fiducial mass resolution (L160N2048) ($p_{\rm med} \sim
0.93$). In the right panel, the impact of varying the \GHeII / \GHeIIavg map
resolution on the \taueffHeII CDF is illustrated for a fixed box size and mass
resolution (L160N2048 model). The \taueffHeII CDF converges effectively when
\GHeII / \GHeIIavg maps are generated at $512^3$ resolution ($p_{\rm med} \sim
0.94$).  The analysis pertains to the redshift range $z = 3.06$ to $z = 3.26$.
}{\label{fig:convergence-test-1}}
%---------------
\InputFigCombine{{Simulation_Resolution_Study_z_2.70}.pdf}{155}%
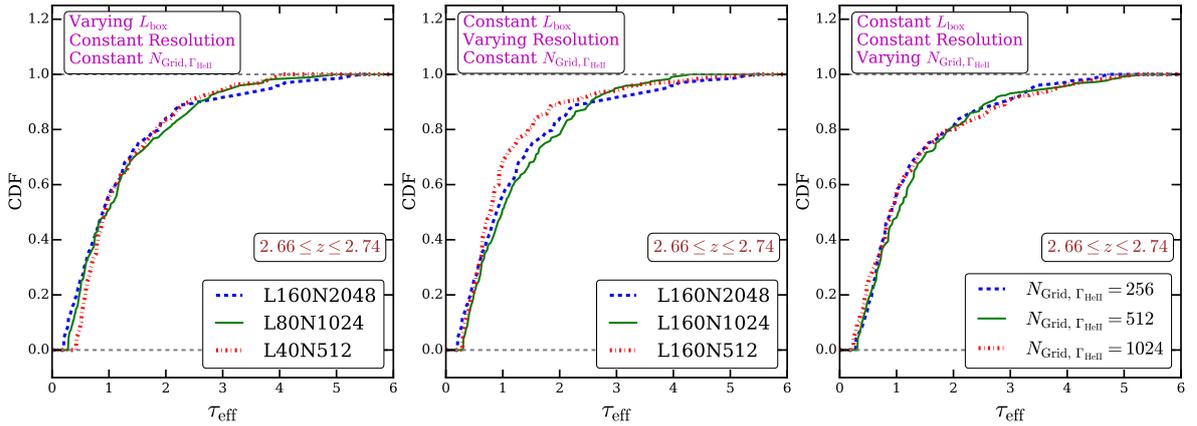
{ Each panel is the same as that of \figref{fig:convergence-test-1} except that the
convergence tests are shown for \zrange{2.66}{2.74}. The convergence test
results are qualitatively similar to that at \zrange{3.06}{3.26}.  The fiducial
model L160N2048 is well converged with respect to box size, mass resolution and
\GHeII / \GHeIIavg map resolution.
}{\label{fig:convergence-test-2}}
%---------------
\InputFigCombine{{Gamma_HeII_Map_Resolution}.pdf}{155}%
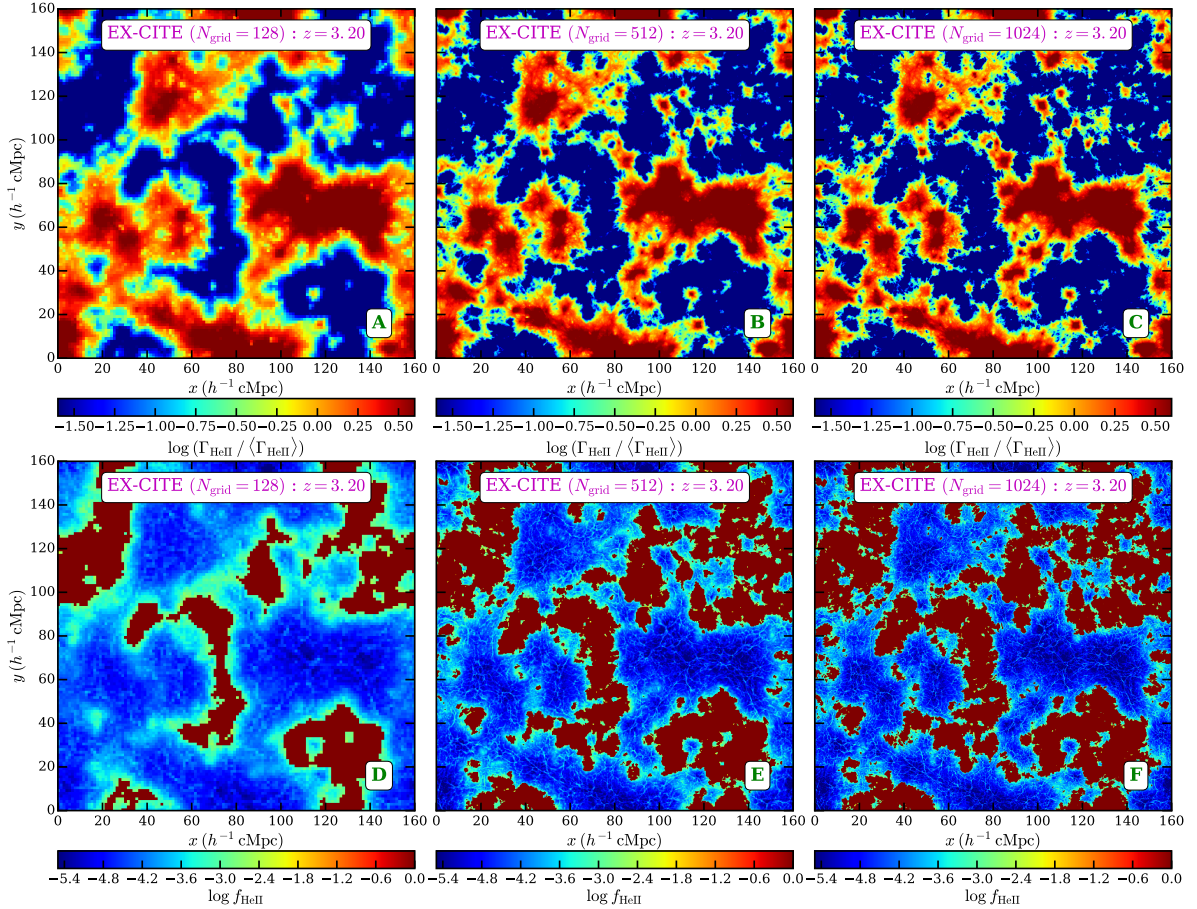
{ Panel A, B, and C show \HeII photo-ionization rate maps ($\Gamma_{\rm HI} /
\langle \Gamma_{\rm HI} \rangle$) generated at resolutions of $128^3$, $512^3$,
and $1024^3$, respectively. Panel D, E, and F show similar maps, but for \HeII
fractions. All simulations are performed with a box of size $L=160 \: h^{-1} \: {\rm cMpc}$ with $N_{\rm particle} = 2048^3$. The color scheme is consistent
across panels for a fair comparison. While large-scale features remain
consistent across resolutions, higher resolutions reveal more substructures,
leading to clumpier distributions on small scales. This highlights the
potential for overestimation of the mean free path at lower resolutions
($<256^3$). Comparing panels B (E) with panels C (F) suggests convergence of
\GHeII / \GHeIIavg and \fHeII maps at resolutions $\geq 512^3$. Most models are
simulated at $512^3$ resolution due to computational constraints. Previous
literature utilized similar methods with $128^3$ resolution, emphasizing the
significance of the enhanced resolution  in improving accuracy for
\GHeIIavg-\LmfpHeII parameter measurements.
}{\label{fig:excite-maps-resolution}}
%---------------

\section{Modeling and Observational uncertainties}
\label{app:parameter-uncertainty}
\figref{fig:parameter-uncertainty-modeling} shows the effect of varying
modeling and the thermal parameters on  \LmfpHeII-\GHeIIavg constraints.
The uncertainty  due to the thermal parameters is dominant and systematic in
nature. The cold model ($T_0-\delta T_0, \gamma + \delta \gamma$) 
predicts systematically higher values of parameters while in the hot model ($T_0+\delta T_0, \gamma
-\delta \gamma$) $1\sigma$ contours are shifted systematically to lower values
of parameters.  The $1\sigma$ uncertainty presented in the previous section
(see also \figref{fig:parameter-constraints-multiple-redshift}) accounts for
the  statistical uncertainty and the thermal  parameters uncertainty.  The
uncertainty in other modeling parameters (i.e., $M_{\rm cutoff}, \beta, \zeta$)
have a marginal effect on the estimated parameters. This is because,  the morphology of the ionizing radiation field is  less sensitive to
these parameters.

\figref{fig:parameter-uncertainty-cosmic-variance} shows the effect of cosmic
variance on the estimated parameters. We compute 1000 $p$ values for each model
while constraining the parameters. In \S \ref{subsec:parameter-constraints}, we
use median $p$ values to constrain the parameters. The scatter in these $p$
values represents the cosmic variance since each $p$ value corresponds to
different skewers.  We assess the effect of cosmic variance on estimated
parameters by using the $16^{\rm th}$ and $84^{\rm th}$ percentile of $p$
values.  We find that the effect of cosmic variance on constrained parameters
is marginal and is within $\sim 2.5$ percent (see
\figref{fig:parameter-uncertainty-cosmic-variance}).

The final source of uncertainty in our estimated parameters is due to 
observational systematics. The observed \taueffHeII are usually subject to
observational systematics due to uncertainty continuum fitting, sky subtraction
and finite \SNR of the observed spectra.  In \S
\ref{subsec:parameter-constraints}, we constrained the parameters using the
observed \taueffHeII without accounting for the uncertainties on those
measurements. \figref{fig:parameter-uncertainty-observations} shows the effect
of the measured \taueffHeII uncertainty on the constraints. The effect of
observational uncertainty is systematic in nature such that the small (large)
values of \taueffHeII leads to larger (smaller) values of the
\LmfpHeII-\GHeIIavg. This is expected because a larger value of \taueffHeII
corresponds to more neutral IGM that would be consistent with models with small
photo-ionization rate and mean free path.
%---------------
\InputFigCombine{{Parameter_Constraints_HeII_thermal_param_uncertainty}.pdf}{155}%
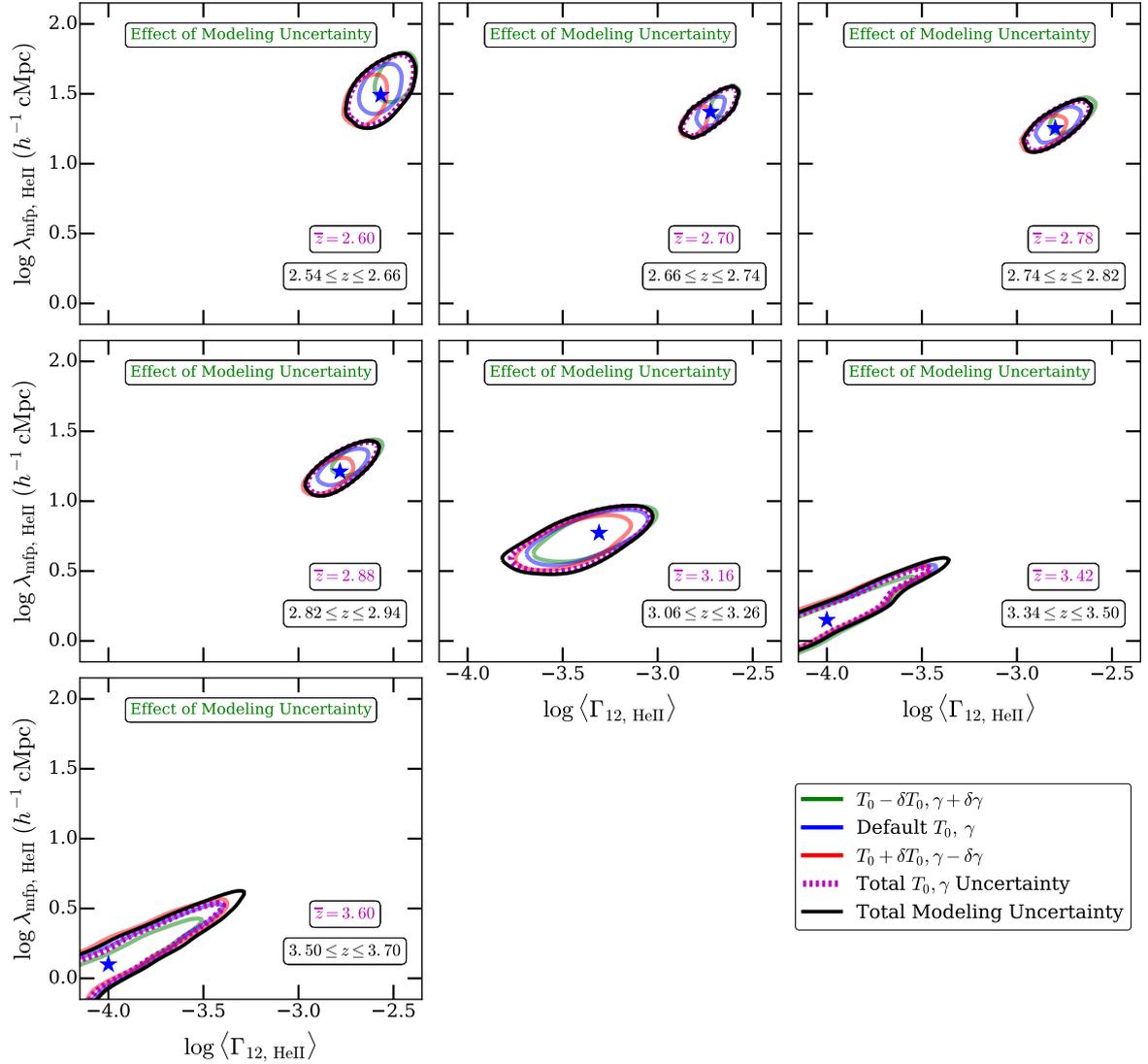
{ Each panel is the same as that in
\figref{fig:parameter-constraints-multiple-redshift} except that the contours are
shown for different thermal parameters of the IGM. The thermal parameters with
$1\sigma$ uncertainty are taken from \cite{gaikwad2021}. The \GHeII
measurements for the hot model i.e., [$T_0 + \delta T_0$,$\gamma-\delta \gamma$] are
systematically lower than those for the best fit model [$T_0,\gamma$]. The \GHeII
measurements for the cold model i.e., [$T_0 - \delta T_0$, $\gamma + \delta \gamma$]
are systematically higher than those for the best fit model. At $z<3$, the variation of \GHeII
with thermal parameters is as expected because \GHeII $\propto \alpha(T)
\propto T^{-0.7}$ for photo-ionization equilibrium. At $z \sim 3.16$, the
sensitivity of \GHeII to thermal parameter decreases. This is because \GHeII
measurements at $z>3$ are more sensitive to reionization topology and density
variation than temperature.  At $z>3.42$, one does not see significant
variation in \GHeII with thermal parameters because \taueffHeII measurements in
this redshift range are dominated by observational systematics (finite \SNR).
Note that other combinations of thermal parameters $T_0 \pm \delta
T_0,\gamma \pm \delta \gamma$ show intermediate shifts in \GHeII.  Thus, the thermal
parameter combination shown in the figure captures the maximum difference in \GHeII.
\textit{The final uncertainty in \GHeII and \LHeII accounts for variation in
thermal parameters.}
}{\label{fig:parameter-uncertainty-modeling}}
%---------------
\InputFigCombine{{Parameter_Constraints_HeII_cosmic_variance_uncertainty}.pdf}{155}%
{
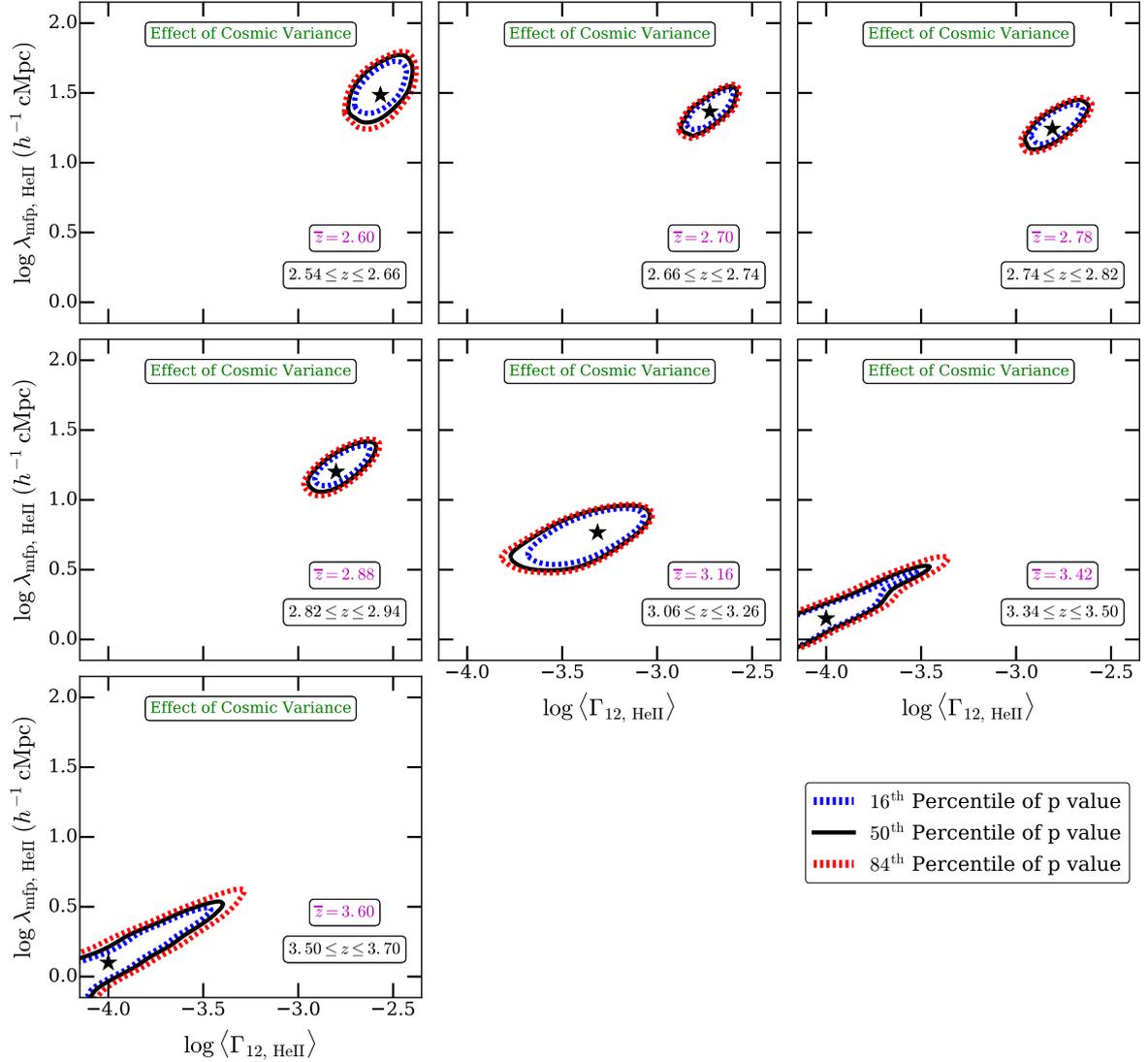
Each panel is the same as that in 
\figref{fig:parameter-constraints-multiple-redshift} except that the contours are
shown for different percentiles of $p$ values. We compare the observed \taueffHeII
CDF for 1000 mocks for each \LmfpHeII-\GHeIIavg. The best fit values and
contours are then calculated using different percentiles of 1000 $p$ values.
The scatter in $p$ values for 1000 mocks corresponds to cosmic variance. By
default we use the $50^{\rm th}$ percentile (median $p$ value, black contours) for
parameter estimation.  The figure shows that if we use the $16^{\rm th}$ (blue
contours) and $84^{\rm th}$ (red contours) percentile of $p$ values, the
contours look very similar. The effect of cosmic variance on \LmfpHeII and
\GHeIIavg is marginal and is within $\sim 2.5$ percent (maximum).  Such a small
difference in the contours is because the $p$ distribution from 1000 mocks is
usually very narrow.  At $z>3.42$, the $p$ distribution is very broad because
of the observational systematics (low \SNR and limited  number of  observed
sightlines).  All the contours shown in this figure also accounts for the
thermal parameter uncertainty.  The final uncertainty in \LmfpHeII and
\GHeIIavg accounts for the uncertainty due to cosmic variance (shown in
\figref{fig:parameter-evolution}).
}{\label{fig:parameter-uncertainty-cosmic-variance}}

%---------------

\InputFigCombine{{Parameter_Constraints_HeII_tau_eff_uncertainty}.pdf}{155}%
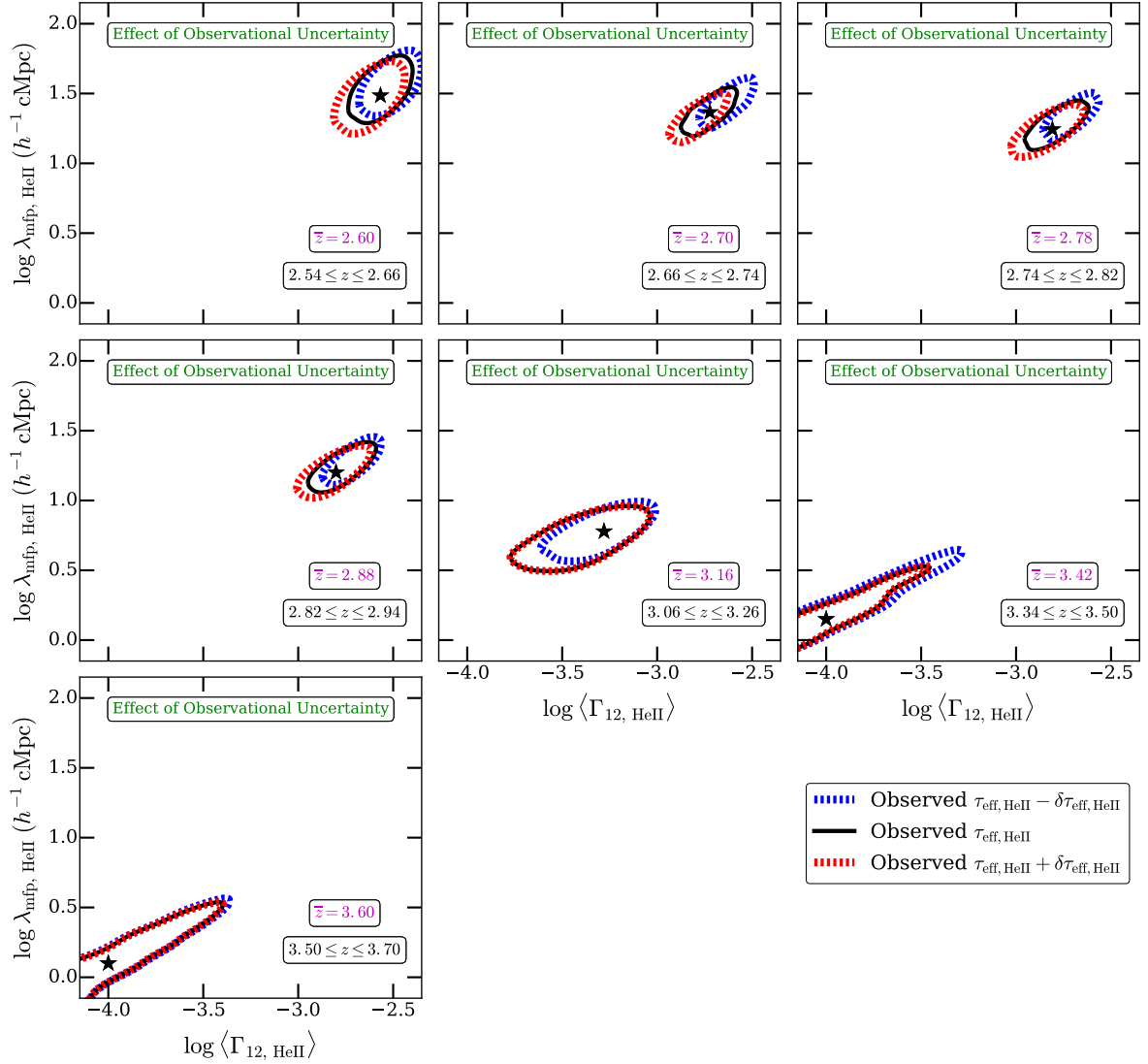
{Each panel is the same as that in 
\figref{fig:parameter-constraints-multiple-redshift} except that the contours show
the effect of observational uncertainty. The blue, black and red contours show
the $1\sigma$ constraints obtained by comparing the observed $\tau_{\rm eff,
HeII} - \delta \tau_{\rm eff, HeII}$, $\tau_{\rm eff, HeII}$, $\tau_{\rm eff,
HeII} + \delta \tau_{\rm eff, HeII}$ respectively with that from simulations.
The observed \taueffHeII and its uncertainty are taken from \cite{worseck2019}
Systematically smaller (higher) values of \taueffHeII result in
systematically larger (smaller) values of \LmfpHeII and \GHeIIavg, respectively.
All the contours shown in this figure also account for the thermal parameter
uncertainty.  The final uncertainty in \LmfpHeII - \GHeIIavg accounts for the
uncertainty due to observational systematics (shown in
\figref{fig:parameter-evolution}).
}{\label{fig:parameter-uncertainty-observations}}
%---------------

%\clearpage
%%\subfile{HeII_excite_appendix_inline}
%%\subfile{HeII_excite_appendix_inline}
%
%
%
%%%%%%%%%%%%%%%%%%%%%%%%%%%%%%%%%%%%%%%%%%%%%%%%%%%
%
%%%%%%%%%%%%%%%%%% APPENDICES %%%%%%%%%%%%%%%%%%%%%
%
%%%%%%%%%%%%%%%%%%%%%%%%%%%%%%%%%%%%%%%%%%%%%%%%%%%
%
%
%% Don't change these lines
%\bsp	% typesetting comment
%\label{lastpage}
\end{document}